\documentclass[a4paper,11pt]{article}

\usepackage[letterpaper,top=2cm,bottom=2cm,left=3cm,right=3cm,marginparwidth=1.75cm]{geometry}

\usepackage{float}
\usepackage{amsmath}
\usepackage{graphicx}
\usepackage[colorlinks=true, allcolors=blue]{hyperref}
\usepackage{caption}
\usepackage{graphicx}
\usepackage{subcaption}
\usepackage{indentfirst}
\usepackage{caption}
\usepackage{cite}

\allowdisplaybreaks[4]

\newcommand{\p}{\partial}
\newcommand{\VEV}[1]{\left\langle #1\right\rangle}
\newcommand{\MeV}{\;\text{MeV}}

\begin{document}

\title{\textbf{A holographic study on QCD phase transition and neutron star properties}}

\author{Xin-Yi Liu,\(^{1,2,3,}\)\footnote{liuxinyi23@mails.ucas.ac.cn} Yue-Liang Wu\(^{1,2,3,4,}\)\footnote{ylwu@itp.ac.cn}  and Zhen Fang\(^{5,6,}\)\footnote{zhenfang@hnu.edu.cn}}
\date{}

\maketitle

\vspace{-10mm}

\begin{center}
    {\it \(^1\) School of Fundamental Physics and Mathematical Sciences, Hangzhou Institute for Advanced Study, UCAS, Hangzhou 310024, China\\
    \(^2\) Institute of Theoretical Physics, Chinese Academy of Sciences, Beijing 100190, China\\
    \(^3\) University of Chinese Academy of Sciences (UCAS), Beijing 100049, China\\
    \(^4\) International Center for Theoretical Physics Asia-Pacific (ICTP-AP), UCAS, Beijing 100190, China\\
    \(^5\) School for Theoretical Physics, School of Physics and Electronics, Hunan University, Changsha 410082, China\\
    \(^6\) Hunan Provincial Key Laboratory of High-Energy Scale Physics and Applications, Hunan University, Changsha 410082, China}
\end{center}


\captionsetup[figure]{name={FIG.},labelsep=period}

\begin{abstract}
We investigate the QCD phase transition and its phase structure within Einstein-Maxwell-Dilaton-scalar system and compare the results with those obtained from the Einstein-Maxwell-Dilaton system. It is shown that both models reproduce behavior consistent with lattice QCD. In particular, the Einstein-Maxwell-Dilaton-scalar system exhibits a first-order phase transition in the pure gauge sector, aligning with predictions from Yang-Mills theory. Based on these models, we construct a holographic model for neutron stars, incorporating leptons to satisfy electric charge neutrality, and examine the cold equation of state, the mass-radius relation, and tidal deformability of neutron stars. It is demonstrated that the Einstein-Maxwell-Dilaton-scalar system enables us to describe neutron star properties that meet current astrophysical constraints.
\end{abstract}

\thispagestyle{empty}
\newpage

\section{Introduction}
Quantum chromodynamics (QCD) is the fundamental theory that describes the strong interaction of quarks and gluons, and it plays a crucial role in astrophysical research. As is well known, the color confinement and the spontaneous chiral symmetry breaking are two basic properties of QCD. At low temperatures, the QCD matter, being in the hadron states, has the property of quark confinement with a nonzero chiral condensate; while at high temperatures, they go from the hadron states into the quark-gluon-plasma states, with the restoration of chiral symmetry \cite{Aoki:2006we, Bazavov:2011nk,Bhattacharya:2014ara}. The research of different phases and phase transitions of QCD matters has always been a hot topic since it is relevant to the formation of matters and the evolution of the universe. However, the study on these topics is remarkably difficult, not only in the theoretical side, because of the nonperturbative nature of QCD at low-energy scales, but also in the experimental side, because fully investigating the QCD phase structure requires extreme conditions, which may include high temperatures, dense environments, and other unique experimental settings, making such experiments particularly challenging.

For research of the QCD matter, the current experiments are mainly using colliders such as the Large Hadron Collider and the Relativistic Heavy Ion Collider, which can create an extremely hot-dense environment similar to the condition of the early universe. However, the experiments on the earth can hardly touch on the issues related to the cold-dense QCD matter. For that, physicists could resort to the compact stars in the universe which supply a natural laboratory for us to study the properties of the cold-dense QCD matter. The neutron star is just one kind of them, which is the remnant after collapse of a massive supergiant star. A newly born neutron star has a relatively high temperature, but after a brief cooling period, its temperature can be approximated as zero relative to the QCD energy scale or to the typical chemical potential of a neutron star. The study of the structure of neutron stars can therefore help us to improve the knowledge of the cold-dense side of the QCD phase diagram. The observation of neutron stars has provided valuable information on their properties, such as mass, radius, and tidal deformability. This, in turn, has allowed for more precise constraints on the neutron star equation of state (EoS), enhancing our understanding of these dense objects.

There have been many theories and models that can be used to probe the phase structure of the QCD matter, such as the lattice QCD \cite{Laermann:2003cv,Fukushima:2013rx}, the chiral effective model and various other theoretical frameworks \cite{Fischer:2009wc,Braun:2009gm,Qin:2010nq,Son:2000xc,Ratti:2005jh,Schaefer:2007pw}. However, almost all the method come with shortages and difficulties due to the nonperturbative nature of low-energy QCD. For the lattice QCD, there was an infamous sign problem at finite chemical potential when considering fermions in the system. Many efforts have been made to develop some extrapolated schemes in order to overcome this problem at finite baryon density, but they can still not span all chemical potentials. Hence, we still need other nonperturbative approaches to tackle the problems relevant to the QCD matter. In the past two decades, the gauge/gravity correspondence \cite{Maldacena:1997re,Gubser:1998bc,Witten:1998qj} has emerged as a powerful tool for addressing various problems, offering a new perspective for exploring this research field \cite{Kruczenski:2003uq, Sakai:2004cn,Sakai:2005yt}. This has led to the development of holographic QCD models, which aim to provide quantitative insights into the nonperturbative properties of QCD, such as hadron spectra, thermodynamics, and the QCD phase structure. Many such models have been developed to date 
\cite{deTeramond:2005su,DaRold:2005mxj,Erlich:2005qh,Karch:2006pv,Cherman:2008eh,Fujita:2009wc,Colangelo:2011sr,Li:2013oda,Policastro:2001yc,Gursoy:2007cb,Gursoy:2007er,Sui:2009xe,Sui:2010ay,Cui:2013xva,Fang:2016dqm,Herzog:2006ra,Barbosa:2024pyn}. Building on this foundation, in this work, we are going to use the holographic method to study the QCD phase transition and also the properties of the neutron star.

In recent years, many studies have employed holographic models to describe neutron stars, including top-down approaches such as the D3/D7 model \cite{Annala:2017tqz, Hoyos:2016zke} and the Witten-Sakai-Sugimoto model \cite{Kovensky:2021kzl, Bartolini:2023wis}. There have also been several bottom-up constructions for modeling neutron stars, such as the V-QCD model \cite{Jokela:2020piw, Demircik:2020jkc} and the Einstein-Maxwell-dilaton (EMD) system \cite{Zhang:2022uin}. For more recent developments on holographic approaches to neutron stars, the reviews \cite{Hoyos:2021uff, Jarvinen:2021jbd} are recommended. In previous works, we explored the QCD phase transition using holographic models \cite{Li:2022erd, Liu:2023pbt}. Based on these studies, we aim to provide a more consistent and realistic holographic description of neutron stars.

In this work, we will pay attention to the phase structure of two-flavor QCD within the Einstein-Maxwell-dilaton-scalar (EMD$\chi$) system \cite{Liu:2023pbt}, which results from combining the EMD system with an improved soft-wall model. As an alternative, we also employ the simplified EMD system to describe the QCD phase transition. The model parameters are determined by comparing with lattice QCD results. We show that both models provide a good description of the QCD equation of state. We then obtain the $\mu-T$ phase diagram for both models, which can display a critical endpoint (CEP), albeit with different values. For the EMD$\chi$ system, we also explore the decoupled EMD system by setting the coupling parameter $\beta = 0$, and find that its EoS is qualitatively consistent with the pure gauge sector of QCD \cite{Boyd:1996bx, Fukushima:2010bq}. Finally, we construct a holographic model of neutron stars based on the above two systems and compute the cold EoS of the neutron star. Consequently, we obtain the mass-radius relation and tidal deformation, which are compared with current astronomical observations.

The paper is organized as follows. In Section~\ref{FormulasPart}, we introduce the EMD and EMD$\chi$ systems with a detailed discussion, and derive the equations of motion (EoM) along with the corresponding boundary conditions. In Section~\ref{QCDPhaseTransitionSection}, we fix the model parameters by fitting to two-flavor lattice QCD results, and then investigate the behavior of the EoS. From this, we extract the properties of the phase transition and obtain the phase diagram. In Section~\ref{NeutronStarConstructMethodSection}, we present the approach used to construct the holographic model of the neutron star. After obtaining the cold EoS, we derive the mass-radius relation and tidal deformability of the neutron star, and compare these results with current constraints. Finally, in Section~\ref{ConclusionSection}, we summarize our work and provide some discussions.

\section{The holographic model for QCD phase transition}\label{FormulasPart}

\subsection{The EMD$\chi$ system}

As shown in our previous work \cite{Liu:2023pbt}, the EMD$\chi$ system can describe the QCD phase transition very well in the two-flavor case. The action of the EMD$\chi$ system can be divided into two parts: \( S = S_G + S_M \). The gravitational background part corresponds to an EMD system, which can be written in the string frame as
\begin{align}\label{act-grav1}
	S_G  = \frac{1}{2\kappa_5^2}\int d^5x\sqrt{-g}e^{-2\phi}\left[R -h(\phi) F_{MN}F^{MN}  +4(\partial\phi)^2 -V(\phi)\right] ,
\end{align}
where \( \kappa_5^2 = 8\pi G_5 \) is the effective Newtonian constant. It is important to note that the EMD system has also been used to describe the QCD system in many works \cite{Gubser:2008yx, Gursoy:2008bu,Zhao:2023gur, Rougemont:2023gfz,Jokela:2024xgz}. In our discussion, we treat the EMD and EMD$\chi$ systems as two distinct models that can be employed to describe the QCD phase transition. The effects of the baryon chemical potential are incorporated through the Abelian gauge field \( A_M \). The dilaton field \( \phi \) breaks conformal symmetry, and can be interpreted roughly as the running coupling of QCD, allowing us to capture the essential behaviors of the QCD phase transition \cite{Gubser:2008yx,Gursoy:2008bu}. Later, we will present the specific form of the dilaton potential \( V(\phi) \), chosen to reproduce the expected thermodynamic properties, along with the gauge kinetic function \( h(\phi) \), which determines the coupling strength of the gauge field \( A_M \).

The flavor part of the EMD$\chi$ action comes from an improved soft-wall AdS/QCD model that is given by
\begin{align}\label{act-flav1}
	S_M  = -\kappa\int d^5x\sqrt{-g}e^{-\phi}\mathrm{Tr}\Big\{|DX|^2 +V_X(X,\phi) +\frac{1}{4g_5^2}(F_L^2+F_R^2)\Big\} ,
\end{align}
where the covariant derivative \( D^M X = \partial^M X - i A_L^M X + i X A_R^M \) and the field strength \( F_{L,R}^{MN} \) is defined as $F_{L,R}^{MN} =\partial^MA_{L,R}^N-\partial^NA_{L,R}^M-i[A_{L,R}^M,A_{L,R}^N]$. The potential for the bulk scalar \( X \) and the dilaton \( \phi \) is given by
\begin{align}\label{VX2}
	V_X(X,\phi) =m_5^2|X|^{2} -\lambda_1\phi |X|^{2} +\lambda_2|X|^{4} ,
\end{align}
where the bulk scalar mass is determined by the mass-dimension relation \( m_5^2 L^2 = \Delta_X (\Delta_X - 4) \), with \( \Delta_X = 3 \) representing the scaling dimension of the dual operator \( \bar{q}_R q_L \) for the scalar field at the boundary \cite{Erlich:2005qh}. We have used this form of the potential in previous works \cite{Li:2022erd,Liu:2023pbt} and found that it successfully captures the QCD phase transition behaviors. In this work, we will use the same potential to construct a cold-dense QCD EoS, which is crucial for modeling neutron stars.

The metric ansatz we adopt for these models is
\begin{align}\label{stringmetric}
	ds^2 &= \frac{L^2 e^{2 A_S(z)}}{z^2} \left(-f(z)dt^2 + dx^i dx^i +\frac{dz^2}{f(z)}\right) ,
\end{align}
where the spacetime has an asymptotic AdS structure in the ultraviolet region (\( z \to 0 \)), and \( L \) is the curvature radius of this spacetime. For simplicity, we set \( L = 1 \) without loss of generality. In the finite temperature case, this metric describes an asymptotic AdS black hole, with \( f(z) \) determined by solving the equations of motion for the system, subject to the boundary condition \( f(z_h) = 0 \) at the event horizon \( z_h \).

The vacuum expectation value (VEV) of the bulk scalar field \( X \) can be written as \( \langle X \rangle = \frac{\chi(z)}{2} I_2 \), where \( I_2 \) is the \( 2 \times 2 \) unit matrix in the two-flavor case, as defined in \cite{Erlich:2005qh}. We neglect the vacuum fluctuations of the meson fields, as these are believed to be negligible compared to the vacuum contribution of the matter fields represented by \( \langle X \rangle \) in the bulk action (\ref{act-flav1}). Thus, the holographic QCD model reduces to an EMD$\chi$ system:
\begin{align}\label{Eintwoscal-str1}
	S &=S_G+S_{\chi}      \nonumber\\
	& =\frac{1}{2\kappa_5^2}\int d^5x\sqrt{-g}e^{-2\phi}\Big[R -h(\phi) F_{MN}F^{MN}  +4(\partial\phi)^2 -V(\phi)-\beta e^{\phi}\Big(\frac{1}{2}(\partial\chi)^2 +V(\chi,\phi)\Big)\Big] ,
\end{align}
where $\beta=16\pi G_5\kappa$ governs the coupling strength between the bulk background and the matter sector. The potential for the scalar VEV \(\chi\) and the dilaton \(\phi\) is given by
\begin{align}\label{Vchi1}
	V(\chi,\phi) & =\mathrm{Tr}\,V_X(\VEV{X},\phi)        \nonumber                          \\
	& =\frac{1}{2}(m_5^2-\lambda_1\phi)\chi^{2} +\frac{\lambda_2}{8} \chi^{4} .
\end{align}

For simplicity, we usually transform to the Einstein frame by employing the following metric ansatz:
\begin{align}\label{einst-metric}
	ds^2 = \frac{L^2 e^{2 A_E(z)}}{z^2} \left(-f(z)dt^2 + dx^i dx^i +\frac{dz^2}{f(z)}\right),
\end{align}
where \(A_E\) relates to the original \(A_S\) through \(A_E = A_S - \frac{2}{3} \phi\). Thus, the action (\ref{Eintwoscal-str1}) in the Einstein frame can be expressed as
\begin{align}\label{Eintwoscal-ef1}
	S &=\frac{1}{2\kappa_5^2}\int d^5x\sqrt{-g_{E}}\Big[R_{E} -w(\phi) F_{MN}F^{MN}   -\frac{4}{3}(\partial\phi)^2-V_E(\phi) -\beta e^{\phi}\Big(\frac{1}{2}(\partial\chi)^2 +V_E(\chi,\phi) \Big)\Big],
\end{align}
where
\begin{align}\label{Vphi-Vchi}
	\begin{split}
		w(\phi) &= e^{\frac{4\phi}{3}} h(\phi),  \\
		V_E(\phi) &= e^{\frac{4\phi}{3}}V(\phi),  \\
		V_E(\chi,\phi) &= e^{\frac{4\phi}{3}}V(\chi,\phi).
	\end{split}
\end{align}
We perform a rescaling $\phi_c= \sqrt{8/3}\,\phi$, to convert the kinetic term of the dilaton $\phi$ into its canonical form. For simplicity, the dilaton potential is taken to have the following simple form as adopted in Ref.~\cite{Gubser:2008yx}:
\begin{align}\label{phi-potent1}
	V_c(\phi_c) =\frac{1}{L^2}\left(-12\cosh\gamma\phi_c +b_2\phi_c^2 +b_4\phi_c^4\right) ,
\end{align}
and set $V_E(\phi) =V_c(\phi_c)$. This ensures that the bulk geometry has an asymptotic AdS structure with $\Lambda = -6$ near the boundary:
\begin{align}\label{phi-potent-uv1}
	V_c(\phi_c\to 0) \simeq \frac{-12}{L^2} +\frac{b_2-6\gamma^2}{L^2}\phi_c^2 +\mathcal{O}(\phi_c^4) .
\end{align}
From mass-dimension relation, we obtain
\begin{align}\label{gamma-b2}
	b_2 =6\gamma^2 +\frac{\Delta(\Delta-4)}{2} ,
\end{align}
where $\Delta$ denotes the scaling dimension of the dual operator of the dilaton field. In principle, when treating the EMD system as a Yang-Mills-like theory, the dilaton \(\phi\) is expected to correspond to the gauge-invariant gluon operator \(\mathrm{Tr}[F^2]\) with \(\Delta = 4\). However, due to the anomaly, the value of \(\Delta\) may deviate from this value \cite{Gubser:2008yx}. Several studies, such as Refs. \cite{Li:2011hp, Cai:2012xh}, have explored the possibility of \(\Delta = 2\), which does not correspond to any local gauge-invariant operator \cite{Gubarev:2000eu, Gubarev:2000nz, Kondo:2001nq}. In fact, the specific value of \(\Delta\) within the Breitenlohner-Freedman bound does not significantly affect the qualitative behavior of the phase transition, and any effects induced by a small deviation in \(\Delta\) can always be absorbed into other model parameters \cite{Li:2022erd}. For the present purposes, we just take \(\Delta = 3\), which has been shown to be a good choice for numerical calculations and is consistent with lattice QCD results \cite{Rougemont:2015wca,Rougemont:2015ona,Cai:2022omk,He:2022amv}.

A particular form of the gauge kinetic function is used,
\begin{align}
	w(\phi) =\frac{1}{4\left(1+c_1\right)}\operatorname{sech} \left(c_{2}\phi_{c}^4\right)+\frac{c_{1}}{4\left(1+c_1\right)}e^{-c_{3}\phi_{c}} ,
\end{align}
which approaches $1/4$ as $\phi_{c}\to 0$ in the UV limit. This form of $w(\phi)$ is motivated in Refs.~\cite{DeWolfe:2010he, DeWolfe:2011ts}, which offers a reliable description of baryon susceptibility at $\mu_{B}=0$.

\subsection{EoM and boundary condition}

We derive the Einstein equation and the EoM for the Abelian gauge field $A_{M}$, the dilaton $\phi$ and the scalar VEV $\chi$ from the above action (\ref{Eintwoscal-ef1}) as follows:
\begin{align}
	& R_{MN} -\frac{1}{2}g_{MN}R +w(\phi)\left(\frac{1}{2} g_{MN} F_{AB}F^{AB}-2F_{MA} F_{N}^{\,\,\,A}\right)      \nonumber              \\
	& +\frac{4}{3}\left(\frac{1}{2}g_{MN}\partial_J\phi\,\partial^J\phi -\partial_M\phi\partial_N\phi\right) +\frac{1}{2}g_{MN}V_E(\phi)       \nonumber \\
	& +\frac{\beta}{2} e^{\phi}\left(\frac{1}{2}g_{MN}\partial_J\chi\,\partial^J\chi -\partial_M\chi\partial_N\chi\right) +\frac{\beta}{2}g_{MN}e^{\phi}V_E(\chi,\phi)  =0 ,   \label{eins-eq0}\\
	& \nabla_{M}\left[w(\phi)F^{MN}\right] =0 ,  \label{EMDS-eomAt}\\
	& \frac{8}{3}\nabla_{M}\nabla^{M}\phi -\partial_{\phi}w(\phi)F_{MN}F^{MN} -\partial_{\phi} V_E(\phi)   \nonumber\\ & -\frac{\beta}{2}e^{\phi} g^{MN} \partial_M\chi \partial_N\chi -\beta\partial_{\phi}\left[e^{\phi}V_{E}(\chi,\phi)\right]   =0 ,   \label{EMDS-phieom}\\
	& \nabla_{M}\left(e^{\phi}\nabla^{M}\chi\right) -e^{\phi}\partial_{\chi}V_{E}(\chi,\phi) =0 .  \label{EMDS-chieom}
\end{align}
It is assumed that only the time-component $A_{t}$ of the Abelian gauge field is nonzero at finite chemical potential, and the bulk fields depend only on the fifth-dimensional coordinate $z$. From the Eqs.~(\ref{eins-eq0})-(\ref{EMDS-chieom}), we obtain the simplified EoMs,  which consist of five independent equations:
\begin{align}
	& f'' +3A_E'f' -\frac{3}{z}f' -4 z^{2} w(\phi) e^{-2 A_{E}} A_{t}^{\prime 2}=0 ,   \label{fz-eom2}\\
	& A_E'' +\frac{2}{z}A_E' -A_E'^2 +\frac{4}{9}\phi'^2 +\frac{\beta}{6}e^{\phi}\chi'^2 =0 , & \label{AE-eom2}\\
	& A_{t}^{\prime\prime}+\left(-\frac{1}{z}+A_E^\prime+\frac{\partial_{\phi} w(\phi)\phi^\prime}{w\left(\phi\right)}\right)A_t^\prime =0,  \label{At-eom2}\\
	& \phi'' +\left(3 A_E' +\frac{f'}{f}-\frac{3}{z}\right)\phi' -\frac{3\beta}{16}e^{\phi}\chi'^2 -\frac{3e^{2A_E} \partial_{\phi}V_E(\phi)}{8z^2f} +\frac{3z^{2} e^{-2 A_{E}} A_{t}^{\prime 2} \partial_{\phi} w(\phi)}{4f}  \notag\\
	& -\frac{3\beta e^{2A_E}\partial_{\phi}\left(e^{\phi}V_E(\chi,\phi)\right)}{8z^{2} f}  =0 ,  \label{dilaton-eom2}\\
	& \chi'' +\left(3A_E' +\phi' +\frac{f'}{f}-\frac{3}{z}\right)\chi' -\frac{e^{2A_E} \partial_{\chi} V_E(\chi,\phi)}{z^2 f}  =0 .  \label{scalarvev-eom2}
\end{align}

The function $f(z)$ and the electrostatic potential $A_{t}(z)$ satisfy the following boundary conditions:
\begin{align}
	f(0) &=1, \qquad   f(z_h)=0 ,  \label{bc-fzAt1}\\
	A_{t}(0) &=\mu_{B} , \qquad 
	A_{t}(z_{h})=0 .   \label{bc-fzAt2}
\end{align}
The UV asymptotic solutions for Eqs.~(\ref{fz-eom2})-(\ref{scalarvev-eom2}) can be obtained as
\begin{align}
	f(z)    & = 1 -f_4 z^4 +\cdots ,     \label{f-uv3}                                                                                                  \\
	A_E(z)  & = -\frac{1}{108}\left(3\beta m_q^2\zeta^2 +8 p_1^2\right)z^2       -\frac{1}{24}\beta p_1 m_q^2\zeta^2 (2\lambda_1+11) z^3 +\cdots ,    \label{AE-uv3}                                                 \\
	A_{t}(z) &=\mu_{B} -\kappa_{5}^{2}n_{B}z^{2} -\frac{4\sqrt{\frac{2}{3}}\kappa_{5}^{2}n_{B}c_1c_3p_1}{3\left(1+c_1\right)}z^3+\cdots , \\
	\phi(z) & = p_1 z +\frac{3}{16} \beta m_q^2\zeta ^2 (\lambda_1+6)z^2 +p_3 z^3       -\left[\frac{1}{48} \beta p_1 m_q^2\zeta ^2 \left(9\lambda_1^2 +111 \lambda_1 +286\right) \right.      \nonumber                    \\
	& \quad \left. -\frac{4}{9} p_1^3 \left(12 b_4-6 \gamma ^4+1\right)\right] z^3\ln z +\cdots ,  \label{phi-uv3}\\
	\chi(z) & =m_q\zeta z +p_1 m_q\zeta(\lambda_1 +5) z^2 +\frac{\sigma}{\zeta} z^3         -\left[\frac{1}{96} m_q^3\zeta^3 \left(\beta  \left(9\lambda_1^2+108 \lambda_1 +308\right)-24 \lambda_2\right) \right.   \nonumber \\  &\quad \left. +\frac{1}{18}p_1^2m_q\zeta \left(9\lambda_1^2 +111 \lambda_1+286\right) \right]z^3\ln z +\cdots ,   \label{chi-uv3}
\end{align}
where $\zeta=\frac{\sqrt{3}}{2 \pi}$ is a normalization constant, $m_q$ represents the current quark mass, $\sigma$ denotes the chiral condensate, \(\mu_B\) is the baryon chemical potential and $n_{B}$ is the baryon number density \cite{DeWolfe:2010he, DeWolfe:2011ts}. From the UV expansions (\ref{f-uv3}) - (\ref{chi-uv3}), we extract two additional conditions:
\begin{align}\label{bc-phichi}
  \phi^{\prime}(0) =p_{1}, \qquad \chi^{\prime}(0) =m_{q}\zeta .
\end{align}
With the boundary conditions (\ref{bc-fzAt1}), (\ref{bc-fzAt2}) and (\ref{bc-phichi}), we can numerically solve Eqs.~(\ref{fz-eom2})-(\ref{scalarvev-eom2}), and the baryon number density $n_B$ can be extracted from the UV asymptotic behavior of $A_t$.

\section{EoS and QCD phase transtion}\label{QCDPhaseTransitionSection}

\subsection{EoS at finite chemical potential}

Solving the system presented above, we can obtain the EoS of the QCD matter according to the holographic dictionary. The temperature $T$ is given by the formula:
\begin{equation}
    T=\frac{|f'(z_h)|}{4\pi} ,
\end{equation}
and the entropy density $s$ is expressed as
\begin{equation}
	s=\frac{2\pi e^{3A_E(z_h)}}{\kappa_5^2 z_h^3} .
\end{equation}
The pressure $p$ can be calculated using the first law of thermodynamics:
\begin{equation}
	dp=sdT+n_Bd\mu_B,
\end{equation}
and the energy density $\varepsilon$ is determined by the thermodynamic relation:
\begin{equation}
	\varepsilon=-p+sT+\mu_Bn_B.
\end{equation}
By using the quantities obtained from the above systems, we can compare the results with lattice QCD data \cite{Karsch:2001vs,Datta:2016ukp}, which allows us to fix the parameters in the present models. The EMD system and EMD$\chi$ system correspond to $\beta=0$ and $\beta=1$, respectively. It is important to note that we can always set the background-matter coupling $\beta=1$ in the EMD$\chi$ system by rescaling the scalar VEV $\chi$ and other parameters. The quark mass in our models is taken to be $m_q=5\MeV$.

\begin{figure}[h!]
\centering 
\includegraphics[width=.48\textwidth]{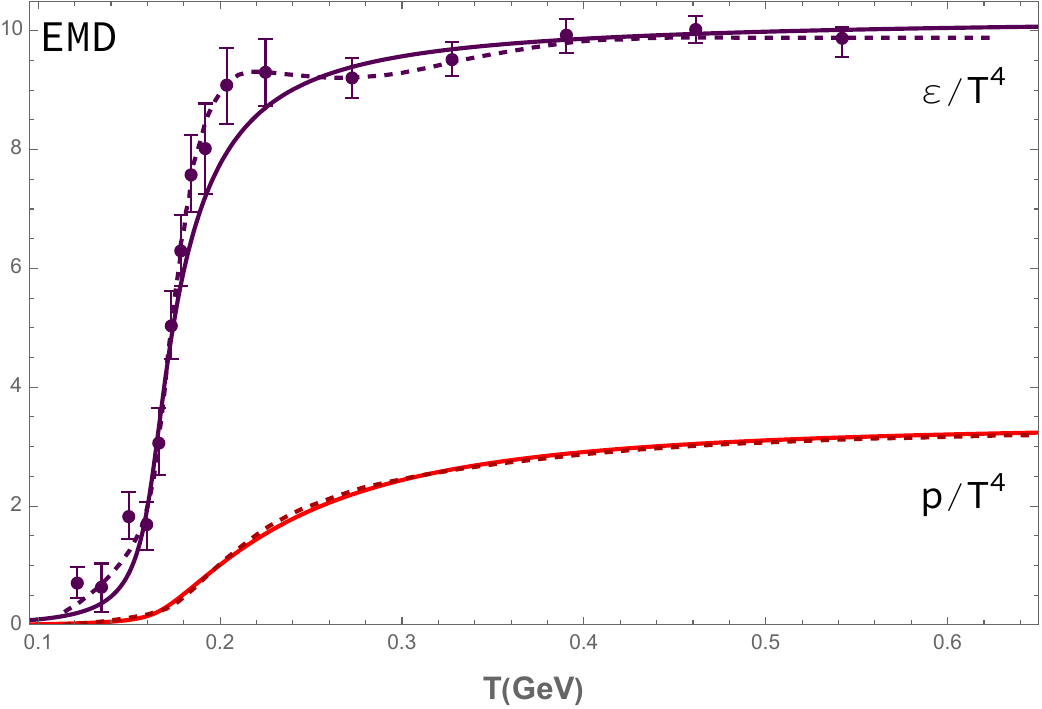}
\hfill
\includegraphics[width=.48\textwidth]{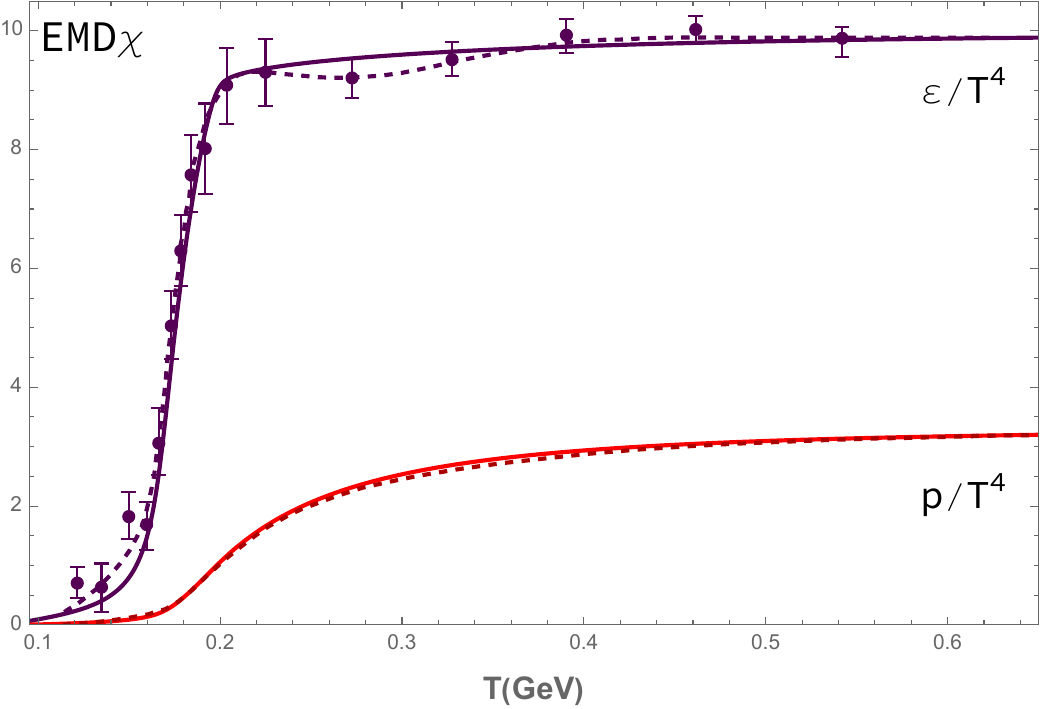}
\caption{A comparison of the results for the scaled energy density \(\varepsilon/T^4\) and pressure \(p/T^4\) with lattice data. The left panel displays the EMD system, while the right panel displays the EMD\(\chi\) system. The solid lines represent the holographic QCD predictions, and the lattice data are shown as dashed lines or points with error bars \cite{Karsch:2001vs}. It is worth noting that the critical temperature \(T_c\) for the lattice data is set to \(T_c = 170 \, \MeV\), which lies within the range discussed in this work.}
\label{ZeroPotentialEoS}
\end{figure}

First, we can turn off the chemical potential part by setting \( \mu_B = 0 \) to fix the parameters that influence the zero potential behavior. For the EMD system, the parameters are determined as: 
\(G_5 = 0.570\), \(\gamma = 0.54\), \(b_4 = -0.125\), and \(p_1 = 0.535\, \text{GeV}\). 
For the EMD\(\chi\) system, the parameters are: 
\(G_5 = 0.582\), \(\gamma = 0.75\), \(b_4 = 0.02\), \(p_1 = 0.487\, \text{GeV}\), \(\lambda_1 = -1\), and \(\lambda_2 = 10\). 
The comparison between the holographic QCD results and the lattice QCD results is illustrated in Fig.~\ref{ZeroPotentialEoS}.

\begin{figure}[h!]
\centering 
\includegraphics[width=.48\textwidth]{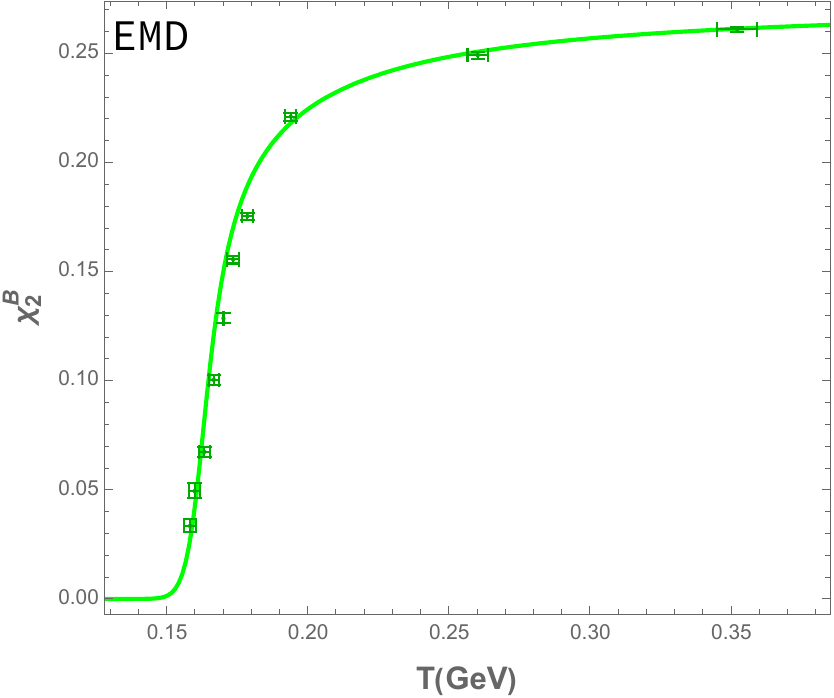}
\hfill
\includegraphics[width=.48\textwidth]{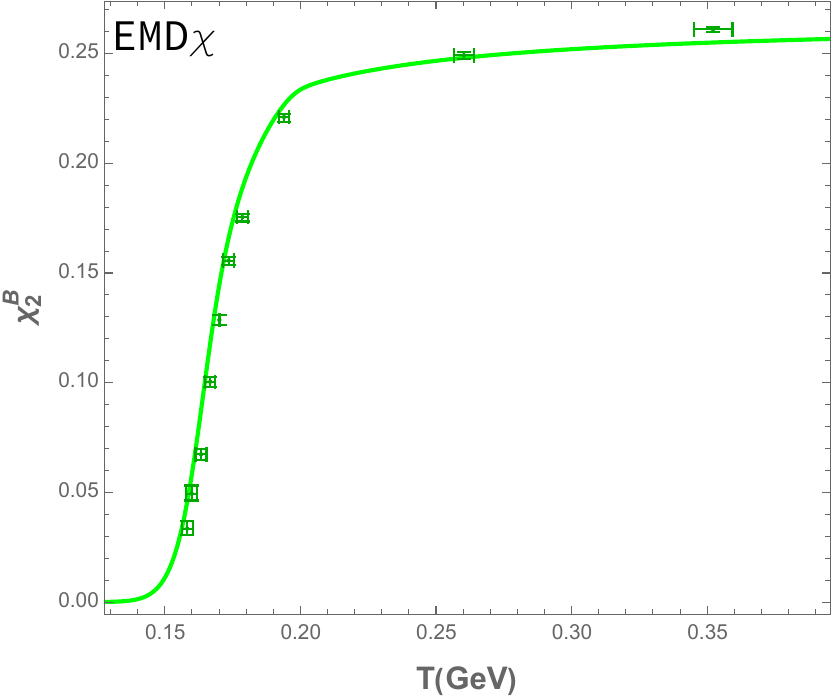}
\caption{A comparison of the results for the baryon number susceptibility $\chi_2^B$  with lattice results. The left panel shows the EMD system, and the right panel shows the EMD$\chi$ system. The soild lines represent the holographic QCD results, while the lattice data are denoted by points with error bars \cite{Datta:2016ukp}. The $T_c$ in these lattice data is also chosen as $T_c=170 \MeV$, the same as the $T_c$ used in Fig.~\ref{ZeroPotentialEoS}.}
\label{ZeroPotentialBNS}
\end{figure}

For non-zero baryon chemical potential $\mu_B$, the remaining parameters to be determined are $c_1,\ c_2$ and $c_3$. We fit the results for the baryon number susceptibility $\chi_2^B$, which is defined as:
\begin{equation}
	\chi^B_2=\frac{\chi_B^2}{T^2}=\frac{\p^2\left(p/T^4\right)}{\p\left(\mu_B/T\right)^2}=\frac{\p\left(n_B/T^3\right)}{\p\left(\mu_B/T\right)}.
\end{equation}
By fitting the lattice results for baryon number susceptibility, we can determine the remaining parameters. For the EMD system: $c_1=1.55,\ c_2=0.026,\ c_3=50;$ for the EMD$\chi$ system: $c_1=1.58,\ c_2=0.018,\ c_3=50.$ The comparisons between the holographic QCD and lattice results are shown in Fig.~\ref{ZeroPotentialBNS}. As seen in the figure, all results are in good agreement with the lattice data.

\begin{figure}[h!]
\centering 
\includegraphics[width=.48\textwidth]{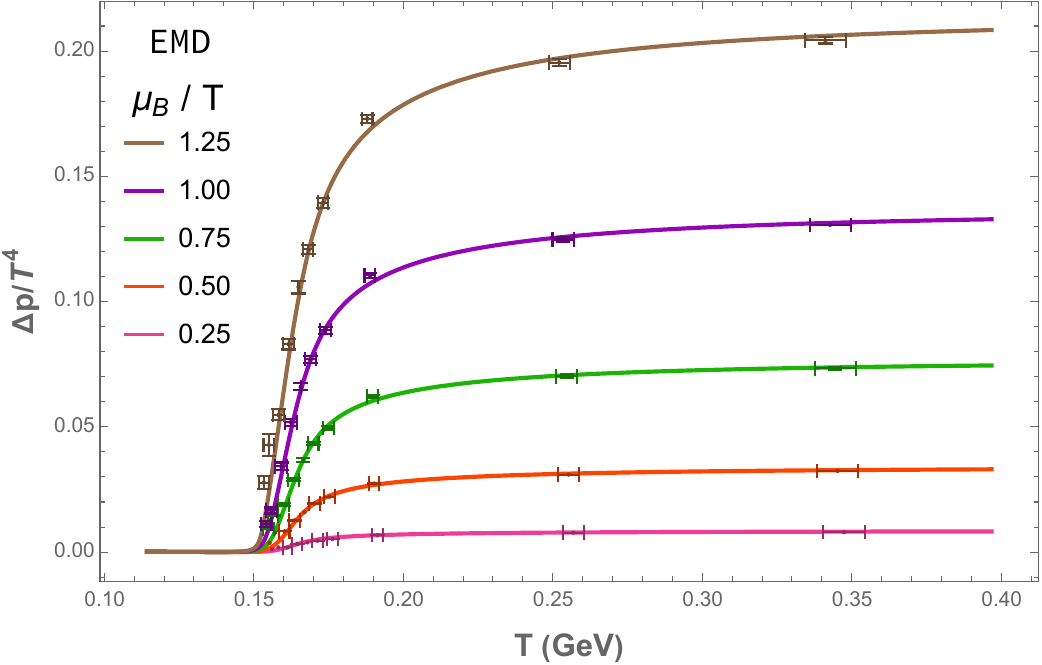}
\hfill
\includegraphics[width=.48\textwidth]{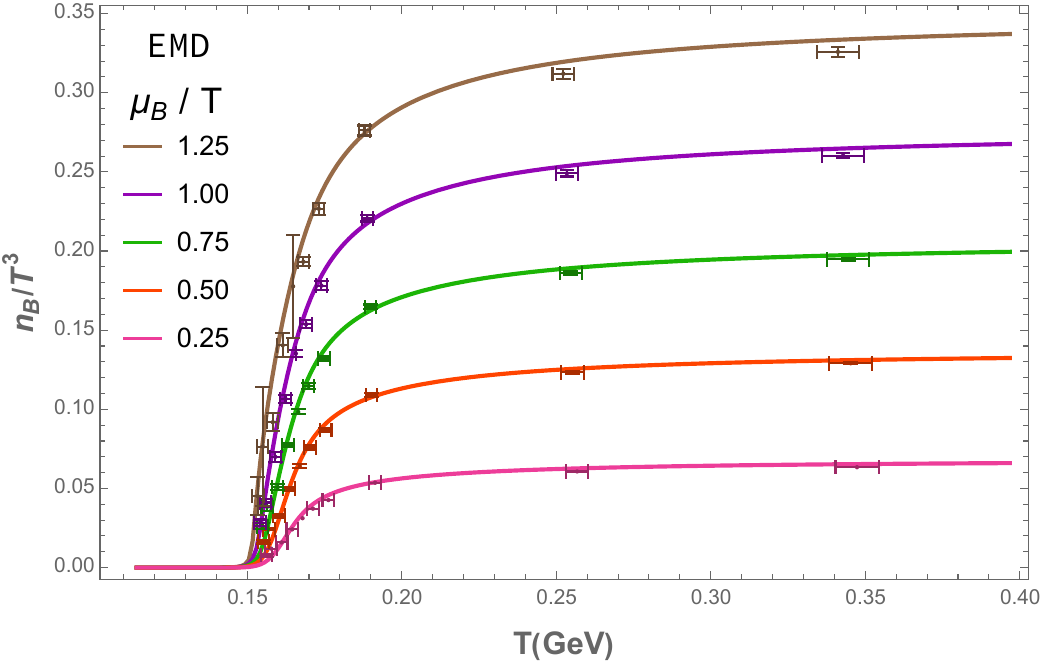}

\includegraphics[width=.48\textwidth]{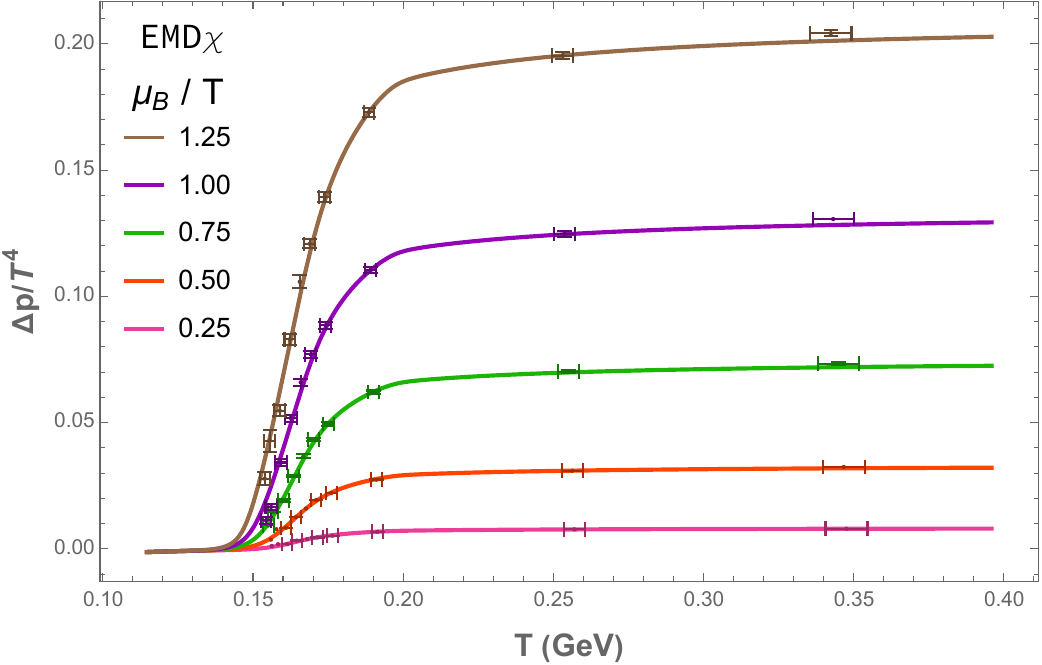}
\hfill
\includegraphics[width=.48\textwidth]{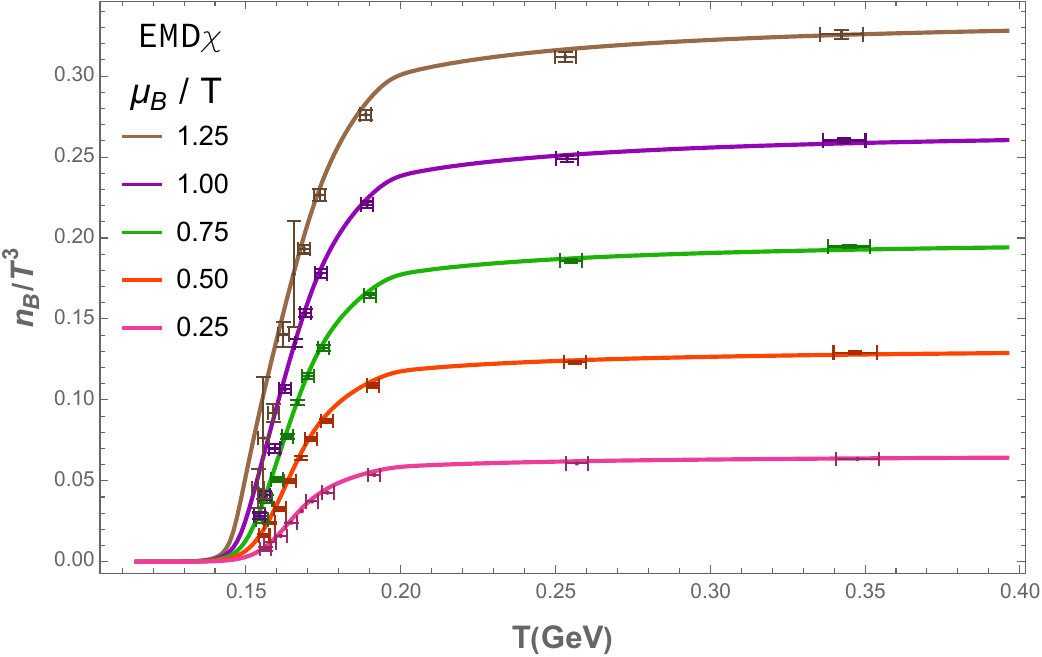}
\caption{A comparison of the model results for the scaled pressure \(\Delta p/T^4\) and baryon number density \(n_B/T^4\) with lattice results for \(\mu_B/T = \left(0.25,\ 0.50,\ 0.75,\ 1.00,\ 1.25\right)\). The upper two panels display the results for the EMD system, while the lower two panels show those for the EMD\(\chi\) system. The finite potential results are taken from Ref.~\cite{Datta:2016ukp}. The lattice \(T_c\) values are chosen as follows: for the EMD system, \(T_c = (167.7,\ 166.7,\ 166.3,\ 165.5,\ 164.7) \MeV\), and for the EMD\(\chi\) system, \(T_c = (167.8,\ 167.4,\ 166.5,\ 165.7,\ 165.3) \MeV\), corresponding to \(\mu_B/T = \left(0.25,\ 0.50,\ 0.75,\ 1.00,\ 1.25\right)\), respectively.}
\label{EoSPotentialOverTCons}
\end{figure}

After determining the parameters in our models, we utilize them to calculate additional EoS results at non-zero chemical potential, such as \(\Delta p/T^4\) and \(n_B/T^3\). These results are then compared with the lattice data from Ref.~\cite{Datta:2016ukp}. 
Here, \(\Delta p\) is defined as \(\Delta p = p(T, \mu_B) - p(T, 0)\). 
The comparison is shown in Fig.~\ref{EoSPotentialOverTCons}.

\subsection{QCD phase diagram}

\begin{figure}[h!]
\centering 
\includegraphics[width=.48\textwidth]{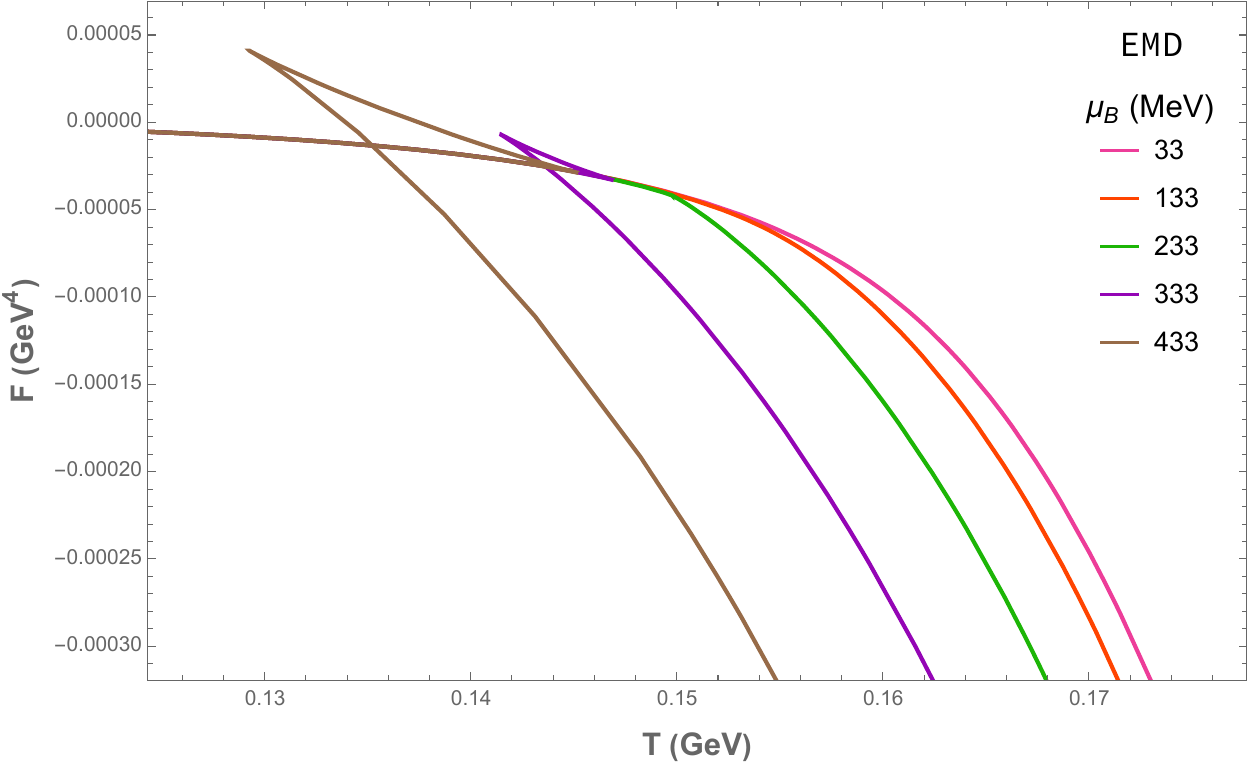}
\hfill
\includegraphics[width=.48\textwidth]{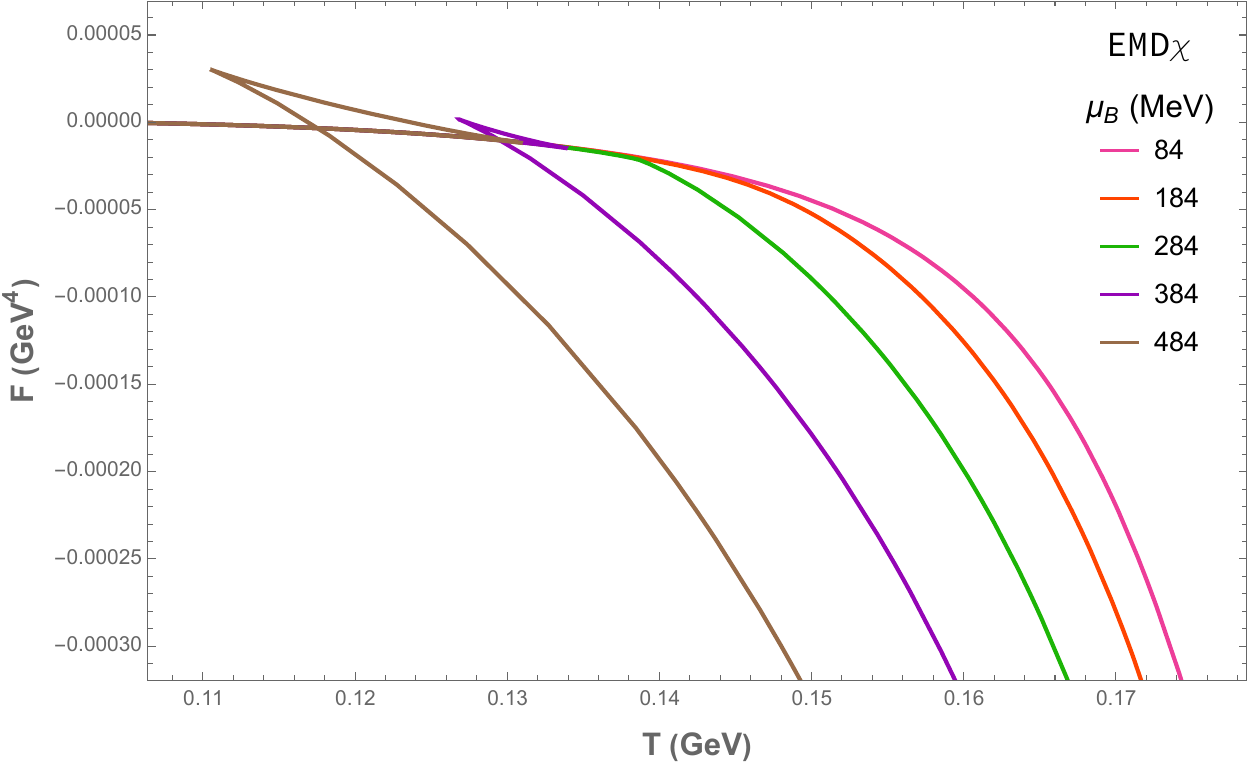}
\caption{The behaviors of free energy $F$ with respect to temperature $T$ at different values of $\mu_B$. The left panel shows the EMD system, and the right panel shows the EMD$\chi$ system. The green line (middle line) represent the CEP potential $\mu_{\rm CEP}$. The transitions for potentials lower than $\mu_{\rm CEP}$ are smooth crossover, while for potentials greater than $\mu_{\rm CEP}$, the transitions are first-order.}
\label{FWithDifferentPotential}
\end{figure}

Now we can calculate the EoS for fixed values of \(\mu_B\) to investigate the behavior of the phase transition. 
We find that at low chemical potential, the free energy density exhibits a smooth crossover. 
As the chemical potential increases, the crossover evolves into a second-order phase transition at the \(\mu_{\rm CEP}\). 
Further increases in potential lead to a first-order phase transition. 
The behavior of the free energy \(F\) as a function of temperature \(T\) is illustrated in Fig.~\ref{FWithDifferentPotential}, where the green lines (middle line) represent second-order phase transitions at the \(\mu_{\rm CEP}\). At lower chemical potentials, the transitions are smooth crossovers, while at higher chemical potentials, they are first-order, characterized by a swallow-tail shape.

For a crossover, there is no unique way to define the transition temperature. However, the maximum increasing point of the baryon number susceptibility \(\chi_2^B\) can be used as an indicator of the crossover temperature. 
As the chemical potential increases, a second-order phase transition occurs at the critical potential \(\mu_{\rm CEP}\). 
In this case, the transition temperature can be determined by analyzing the behavior of the free energy \(F\) as a function of temperature \(T\). 
As the potential increases further, a first-order phase transition takes place, characterized by the emergence of a swallow-tail shape in the free energy density, as shown in Fig.~\ref{FWithDifferentPotential}. 
The temperature at the intersection point of the free energy curves corresponds to the first-order phase transition temperature. 
Using this method, we can determine the temperature at the CEP, \(T_{\rm CEP}\), by identifying the temperature at which the swallow-tail shape first appears, marking the occurrence of the second-order phase transition.

\begin{figure}[h!]
\centering 
\includegraphics[width=.6\textwidth]{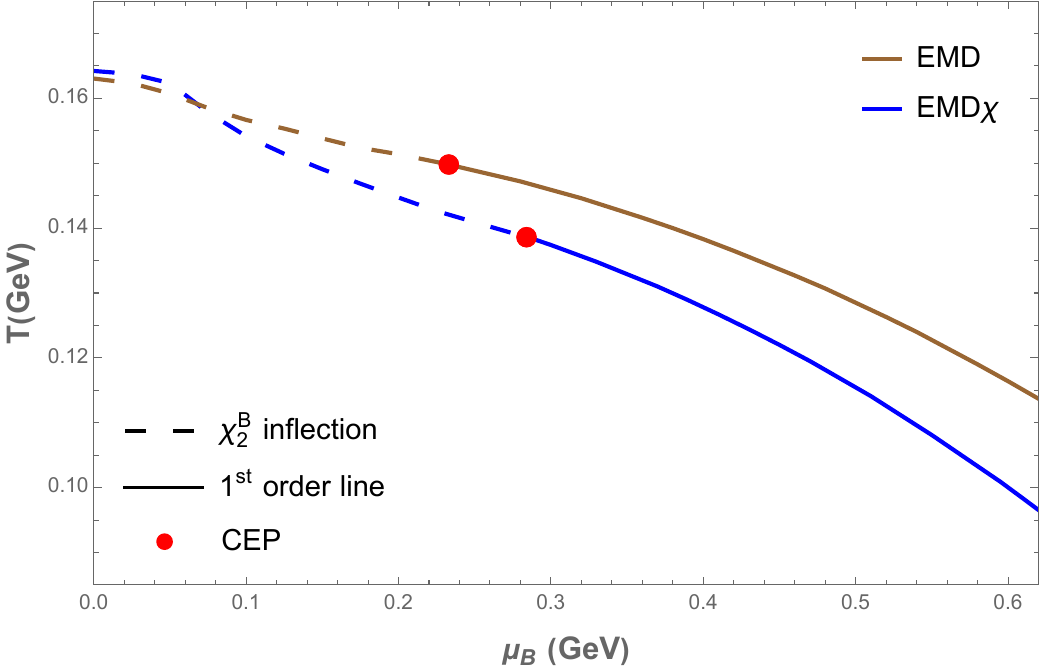}
\caption{The QCD phase diagram from two systems. The brown line represents the EMD system, and the blue line represents the EMD$\chi$ system.  The dashed lines indicate the maximum increasing point of the baryon number susceptibility, while the solid lines denote the first-order phase transitions. The CEP is marked by a red point.}
\label{PhaseDiagram}
\end{figure}

By investigating the EoS behaviors at various fixed values of chemical potential, we can construct the QCD phase diagrams for our models, which have been shown in Fig.~\ref{PhaseDiagram}.
The locations of the CEP in our systems are as follows: for the EMD system, \(\left(\mu_{\rm CEP}, T_{\rm CEP}\right) = \left(233 \, \MeV, 149.8 \, \MeV\right)\), and for the EMD\(\chi\) system, \(\left(\mu_{\rm CEP}, T_{\rm CEP}\right) = \left(284 \, \MeV, 138.6 \, \MeV\right)\). Below \(\mu_{\rm CEP}\), the system undergoes a smooth crossover, while above \(\mu_{\rm CEP}\), a first-order phase transition occurs. The CEP serves as the connection point between these two types of phase transitions. We compare our findings with other studies and observed that the locations of the CEP in our models are close to those reported in Ref.~\cite{Gao:2016qkh}, particularly for the EMD\(\chi\) system. When comparing with other works \cite{Sasaki:2010jz,Fu:2019hdw,Gao:2020qsj,Zhao:2023gur}, our results fall between theirs, though significant differences in CEP predictions still exist for different models.

\subsection{Pure gauge sector}

\begin{figure}[h!]
\centering 
\includegraphics[width=.46\textwidth]{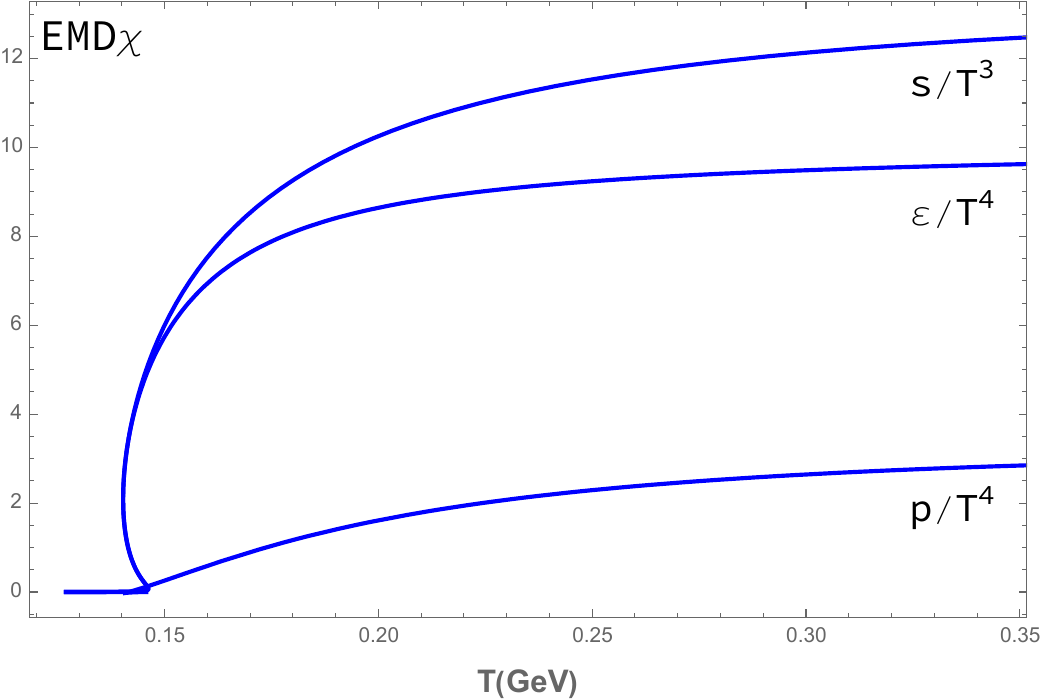}
\hfill
\includegraphics[width=.49\textwidth]{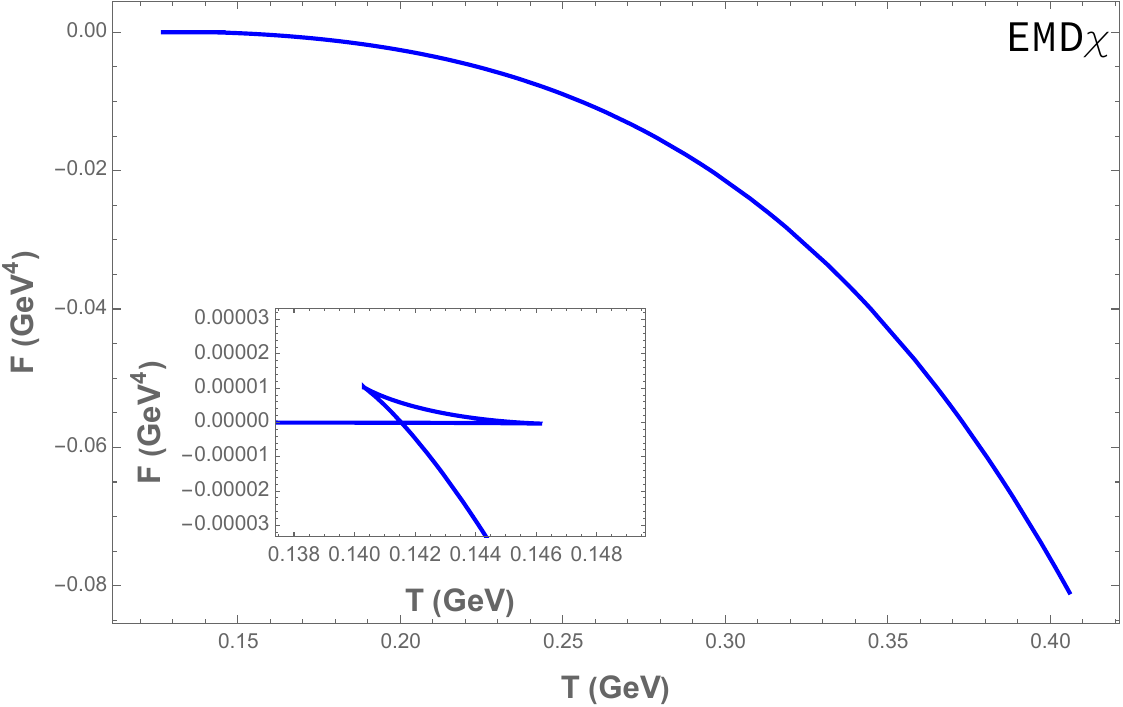}
\caption{The behavior of the scaled entropy density \(s/T^3\), energy density \(\varepsilon/T^4\), and pressure \(p/T^4\) as functions of temperature \(T\) are shown in the left panel, while the free energy \(F\) is presented in the right panel. A distinct swallow-tail shape, characteristic of a first-order phase transition, is observed.}
\label{PureGauge}
\end{figure}

In our EMD\(\chi\) system, we set \(\beta = 1\) to describe the QCD system. 
To decouple the matter part from the pure gauge sector, we set \(\beta = 0\), while keeping all other parameters identical to those in the EMD\(\chi\) system. Repeating the numerical calculations, we can determine the behaviors of the EoS under this setup. 
The numerical results are shown in Fig.~\ref{PureGauge}, where we compute the scaled entropy density \(s/T^3\), energy density \(\varepsilon/T^4\), pressure \(p/T^4\), and free energy \(F\), all at zero chemical potential, \(\mu_B = 0 \).
We observe a swallow-tail shape in the results, indicating a first-order phase transition. This outcome is consistent with expectations and aligns with lattice simulations for pure Yang-Mills theory \cite{Boyd:1996bx, Fukushima:2010bq}. From the right panel of Fig.~\ref{PureGauge}, we extract the critical temperature which is approximately \(T_c \simeq 141.5 \, \MeV\).

\section{Holographic neutron star} \label{NeutronStarConstructMethodSection}
\subsection{The core structure} \label{NeutronStarConstructMethodSection-core}

As mentioned earlier, our models can be employed to describe the QCD system. 
We now apply them to obtain the cold EoS, which is used to construct the baryonic component of the neutron star. To distinguish the baryonic component from the subsequent leptonic component, we introduce the subscript ``B'' to denote the baryonic component. The temperature and entropy density are given by the following formulas:
\begin{align}\label{temperentro}
	T_B=\frac{|f'(z_h)|}{4\pi} , \qquad  s_B =\frac{2\pi e^{3A_E(z_h)}}{\kappa_5^2 z_h^3} .
\end{align}
The expressions for the pressure $p_B$ and energy density $\varepsilon_B$ are:
\begin{align}
	&dp_B=s_B dT_B +n_{B}d\mu_{B}  \label{pTdiffrelat},\\
	&\varepsilon_B=-p_B+s_BT_B+\mu_B n_B \label{EnergyDensity}.
\end{align}

For a more realistic neutron star model, it is necessary to include leptons. 
In this work, we incorporate a non-interacting free lepton gas, considering only electrons with a mass of \(m_e = 511\, {\rm keV}\), to ensure charge neutrality. In the core, we assume local charge neutrality (where ``local'' refers to the condition relative to the global charge neutrality used when constructing the neutron star crust):
\begin{equation}\label{LocalChargeNeutrality}
  n_P =n_L,
\end{equation}
where $n_P$ and $n_L$ denote the proton and lepton number densities, respectively. Strictly speaking, the isospin density $n_{I}$ should be introduced when we talk about the proton number density. However, as a first attempt, we do not consider $n_{I}$ in this work, but only take $n_P/n_{B}$ as a free parameter. The lepton number density, pressure and energy density are related to the lepton chemical potential by the following zero-temperature Fermi gas expression \cite{Schmitt:2010pn}:
\begin{align}
n\left(m,\mu\right) &\equiv \Theta\left(\mu-m\right) \frac{\left(\mu^2-m^2\right)^{3/2}}{3\pi^2}\label{LeptonChemicalPotential},\\
	p\left(m,\mu\right)&\equiv\frac{\Theta\left(\mu-m\right)}{24\pi^2}\left[\left(2\mu^2-5m^2\right)\mu\sqrt{\mu^2-m^2}+3m^4 \ln \frac{\sqrt{\mu^2-m^2}+\mu}{m}\right]\label{LeptonFreeEnergy},\\
 \varepsilon\left(m,\mu\right)&\equiv\frac{\Theta\left(\mu-m\right)}{8\pi^2}\left[\left(2\mu^2-m^2\right)\mu\sqrt{\mu^2-m^2}-m^4\ln\frac{\sqrt{\mu^2-m^2}+\mu}{m}\right]\label{LeptonEnergy}.
\end{align}
The total pressure and energy density in the core of neutron star are obtained by summing the contributions from both the baryonic and leptonic components:
\begin{align}
	p_{core} &=p_L+p_B,\\	\varepsilon_{core}&=\varepsilon_L+\varepsilon_B.
\end{align}

\subsection{The crust structure} \label{NeutronStarConstructMethodSection-crust}

As discussed in Ref.~\cite{Kovensky:2021kzl}, numerous previous holographic studies have utilized traditional methods to construct the crust, relying on well-established nuclear physics at low densities. While this is a reasonable approach, significant uncertainties remain regarding the crust-core transition. In light of this, we apply the holographic framework consistently across all density regions of the neutron star, allowing the crust-core transition to be determined dynamically.

To construct the crust of the neutron star, we introduce a mixed phase consisting of two coexisting components \cite{Kovensky:2021kzl}: the nuclear matter phase (which also exists in the core), composed of both baryonic and leptonic matter treated as a single phase, and the lepton gas phase, which appears in the crust. We define \(\xi \in [0,1]\) as the volume fraction of the leptonic phase, with the nuclear phase occupying a fraction of \(1 - \xi\). The conditions for the mixed phase are specified as follows:
\begin{align}
    p_B &= 0, \label{Core-Crust transition point}\\
    \left(1 - \xi\right)\left(n_P - n_L\right) - \xi n_L &= 0. \label{GlobalChargeNeutrality}
\end{align}
The first equation arises from the requirement that the pressure of the nuclear matter phase, \(p_B + p_L\), is equal to the pressure of the leptonic phase, \(p_L\), which implies that the baryonic pressure component \(p_B\) must be zero in the crust. The second equation enforces the global charge neutrality condition. Additionally, Eq. (\ref{Core-Crust transition point}) provides a criterion for determining the crust-core transition point.

To solve for the quantities in the crust, we first set \(n_B\) in the crust equal to its value at the point where \(p_B = 0\) in the core EoS, which is obtained from the previous holographic procedure. 
Then, we obtain a relation between \(n_L\) and \(\xi\) using Eq.~ (\ref{GlobalChargeNeutrality}) once $n_{P}$ were determined. It should be noted that \(n_B\) remains constant throughout the crust, but its average value, \(\left<n_B\right> = \left(1 - \xi\right)n_B\), decreases and approaches zero as the system moves toward vacuum. The pressure and energy density in the crust can be expressed as
\begin{align}
   p &=p_L,\\
    \varepsilon &=(1-\xi) \left(\varepsilon_B +\varepsilon_L\right) +\xi\varepsilon_L \label{TotalEnergyInTheCrust},
\end{align}
where \(p_L\) and \(\varepsilon_L\) are obtained from Eqs.~(\ref{LeptonFreeEnergy}) - (\ref{LeptonEnergy}), and \(\varepsilon_B\) is evaluated at the point where \(p_B = 0\) in the core EoS, similar to \(n_B\).

Now that we have a basic construction of the neutron star, we recognize that the current EoS does not incorporate the geometric structure of the mixed-phase crust, for which the surface and Coulomb effects should be considered. As in Ref.~\cite{Kovensky:2021kzl}, we also employ the Wigner-Seitz approximation, where the geometric details are reflected in the shape of the unit cell.
The contribution to the free energy from the surface and Coulomb effects is given by
\begin{align}\label{SurfaceAndCoulombTerm}
    \Delta F =\frac{3}{2}\left[e\left(n_P-n_L\right)-e\left(-n_L\right)\right]^{2/3} \Sigma^{2/3}\left(1-\xi\right)\left[d^2f_d\left(1-\xi\right)\right]^{1/3},
\end{align}
where \(e=\sqrt{4\pi\alpha}\simeq0.3\), and we choose \(d=3\), which corresponds to the bubble geometric structure in the Wigner-Seitz approximation. In this case \cite{Schmitt:2020tac},
\begin{align}
    f_3\left(\xi\right)\equiv \frac{2+\xi-3\xi^{1/3}}{5} 
\end{align}
Note that in this scenario the nuclear matter is immersed within the lepton gas which is the phase in the outer region of the Wigner-Seitz cell.
The parameter \(\Sigma\) represents the surface tension, treated as an external parameter in the step-like approximation of the interface profiles. In principle, \(\Sigma\) is a dynamic variable that depends on baryon and isospin densities, but for simplicity, we just keep \(\Sigma\) as a constant within the neutron star crust. The final pressure and energy density of the mixed-phase crust are given by
\begin{align}
    &p_{\text{crust}}=p_L-\Delta F\label{CrustTotalPressureWithSurface},\\
    &\varepsilon_{\text{crust}}=(1-\xi) \left(\varepsilon_B +\varepsilon_L\right) +\xi\varepsilon_L+\Delta F\label{CrustTotalEnergyWithSurface}.
\end{align}

\subsection{EoS of neutron stars}\label{HoloCal-1}

Now we proceed by applying the above procedure to compute the EoS of the neutron star. We carry out our calculation at the lowest achievable temperatures: \(T = 0.0176 \,\text{MeV}\) for the EMD system, and \(T = 0.0233 \, \text{MeV}\) for the EMD\(\chi\) system. It is worth noting that the slight temperature difference between 0.0176 MeV and 0.0233 MeV has a negligible effect on the results.

In our model, there is flexibility in selecting the zero point of the baryon pressure \(p_B\). To achieve a stiffer EoS for the neutron star within the permissible range, we set \(p_B = 0\) at \(\mu_B = 1.0205 \, \text{GeV}\) for the EMD system and \(\mu_B = 0.9685 \, \text{GeV}\) for the EMD\(\chi\) system. As discussed above, we treat the ratio \(n_P/n_B\) as a free parameter, exploring different values of \(n_P/n_B\) and then comparing the results with observational constraints. For each selected ratio, we calculate \(n_P\) and use Eq.~(\ref{LocalChargeNeutrality}) to determine \(n_L\). Subsequently, we apply Eq.~(\ref{LeptonChemicalPotential}) to solve for the lepton chemical potential, and use Eqs.~(\ref{LeptonFreeEnergy}) - (\ref{LeptonEnergy}) to obtain the pressure and energy of the leptons in the core.

\begin{figure}[h!]
\centering 
\includegraphics[width=.48\textwidth]{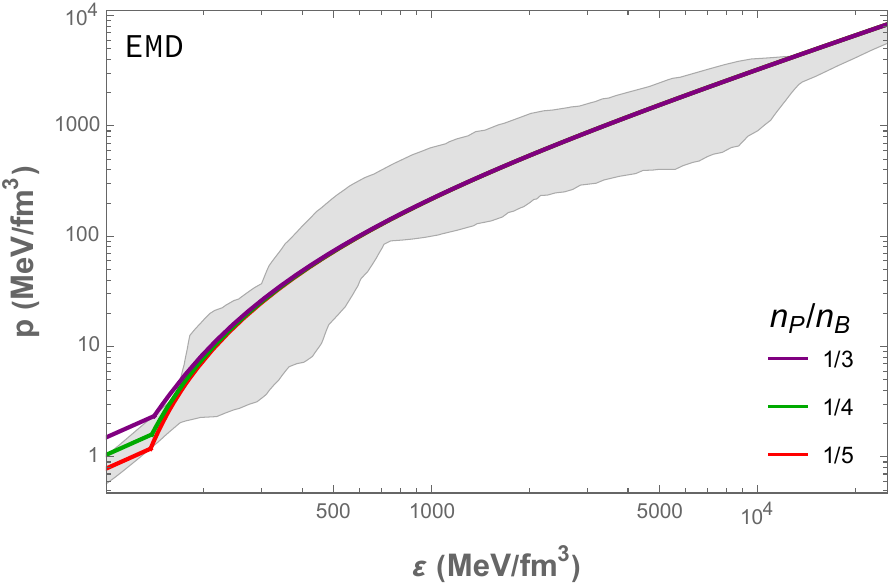}
\hfill
\includegraphics[width=.48\textwidth]{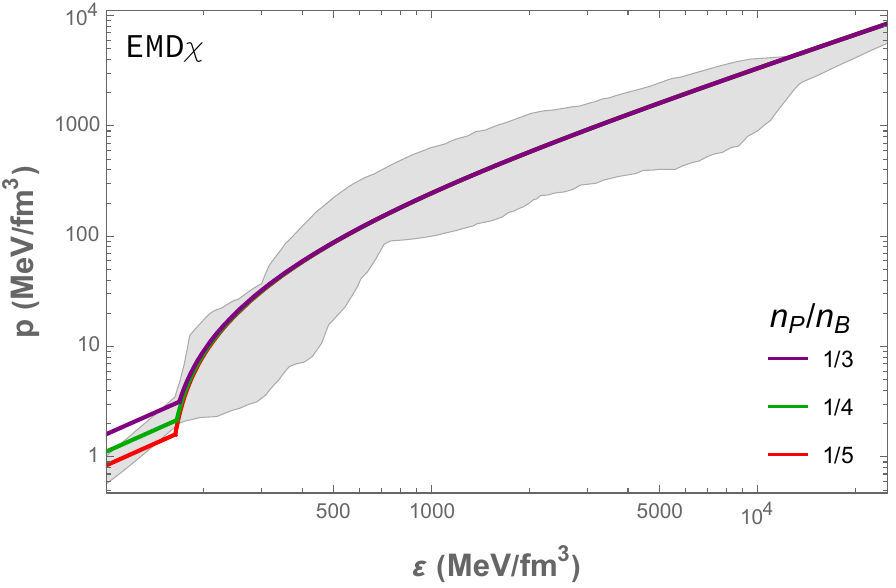}
\caption{Equations of state of the neutron star for three different ratios, \(n_P/n_B =1/3,\,1/4,\,1/5\), represented by purple, green, and red lines, respectively. The left panel represents the EMD system, while the right panel corresponds to the EMD$\chi$ system. The gray band, taken from Ref.~\cite{Annala:2019puf}, indicates the allowed range for the cold EoS.}
\label{NeutronStarPEFig}
\end{figure}

At this stage, we have acquired all necessary information for the core, enabling us to determine the crust-core transition point using Eq.~(\ref{Core-Crust transition point}). Following the specified procedure for the crust, we can then construct the complete EoS of the neutron star. Initially, we set \(\Sigma = 0\) to examine the effects of different \(n_P/n_B\) ratios on the EoS. We consider three different ratios, \(n_P/n_B = 1/3, \, 1/4, \, 1/5\), as illustrated in Fig.~\ref{NeutronStarPEFig}.
The gray band constraint for the EoS is taken from Ref.~\cite{Annala:2019puf}. The low-energy density region is derived from chiral effective theory \cite{Hebeler:2013nza}, while the high-energy density region is determined by NNLO perturbative QCD calculations \cite{Kurkela:2009gj}. The intermediate region is constructed using a new speed-of-sound interpolation method that incorporates astrophysical constraints. It is important to emphasize that significant uncertainties remain between the regions of validity of chiral effective theory and perturbative QCD, with various constraints derived through different methods \cite{Annala:2017llu,Altiparmak:2022bke}.

As shown in Fig.~\ref{NeutronStarPEFig}, the EoS curves for different values of \( n_P / n_B \) overlap in the core region but diverge in the crust. This indicates that variations in \( n_P / n_B \) have a minimal effect on the core properties of the neutron star, but significantly influence the EoS behavior in the crust. By comparing with the allowed region, we will adopt \( n_P / n_B = 1/4 \) for the subsequent analysis. After incorporating the contributions from surface and Coulomb effects, we will derive the final neutron star EoS, and then we will use the \( p \)-\( \varepsilon \) relation to compute quantities that can be compared with astronomical observations, such as the mass-radius (M-R) relation and the tidal deformability \( \Lambda \).

\subsection{TOV equation and Tidal deformability}

The Tolman-Oppenheimer-Volkoff (TOV) equation for a spherically symmetric, isotropic object in static gravitational equilibrium is expressed as \cite{Tolman:1934za, Tolman:1939jz, Oppenheimer:1939ne}:
\begin{align}\label{TOVEq}
	\frac{dp(r)}{dr} + \left(p(r) + \varepsilon(r)\right)\frac{G\left(m(r) + 4\pi r^3 p(r)\right)}{r^2\left(1 - 2\frac{Gm(r)}{r}\right)} = 0,
\end{align}
where \( r \) is the radial coordinate, \( G \) is the Newtonian gravitational constant, and \( p(r) \), \( \varepsilon(r) \), and \( m(r) \) represent the pressure, energy density, and total mass enclosed within radius \( r \), respectively. The mass function \( m(r) \) satisfies the following differential equation:
\begin{align}\label{MassEq}
  \frac{dm(r)}{dr} = 4\pi r^2 \varepsilon(r).
\end{align}
With the EoS as input, the M-R relation of the neutron star can be obtained by solving Eqs.~(\ref{TOVEq}) - (\ref{MassEq}) with the boundary conditions:
\begin{align}
	m(0) &= 0, \label{bc-NS1} \\
	p(0) &= p_c, \label{bc-NS2}
\end{align}
where \( p_c \) is the central pressure of the neutron star, provided as an input parameter. Once the functions \( p(r) \), \( \varepsilon(r) \), and \( m(r) \) are determined, the radius \( R \) of the neutron star is identified by locating the point where \( p(R) = 0 \), and the total mass of the neutron star is given by \( M \equiv m(R) \).

The tidal deformability is calculated by solving the following differential equation \cite{Hinderer:2007mb,Hinderer:2009ca,Postnikov:2010yn}:
\begin{align}\label{TidalDefEq}
	&r\frac{dy(r)}{dr} + y^2(r) + \frac{4\pi G r^2 \left( 5\varepsilon(r) + 9p(r) + \frac{\varepsilon(r) + p(r)}{c_s^2} \right) - 6}{1 - \frac{2Gm(r)}{r}} \nonumber \\
	&+ \frac{y(r)\left[ 1 - 4\pi G r^2 (\varepsilon(r) - p(r)) \right]}{1 - \frac{2Gm(r)}{r}} - \frac{4G^2 \left( m(r) + 4\pi p(r) r^3 \right)^2}{r^2 \left( 1 - \frac{2Gm(r)}{r} \right)^2} = 0,
\end{align}
where \( y(r) \) represents the metric perturbation, with the boundary condition \( y(0) = 2 \), derived from the asymptotic solution of Eq.~(\ref{TidalDefEq}) as \( r \to 0 \). Using these quantities, we can then calculate the tidal Love number:
\begin{align} 
k_2= & \frac{8 c^5}{5}(1-2 c)^2\left[2-y_R+2 c\left(y_R-1\right)\right]  \times\left\{2 c\left[6-3 y_R+3 c\left(5 y_R-8\right)\right]\right. \nonumber\\ & +4 c^3\left[13-11 y_R+c\left(3 y_R-2\right)+2 c^2\left(1+y_R\right)\right]\nonumber \\ & \left.+3(1-2 c)^2\left[2-y_R+2 c\left(y_R-1\right)\right] \ln (1-2 c)\right\}^{-1},
\end{align}
where \( y_R \equiv y(R) \), and \( c = GM/R \) denotes the compactness of the star. Finally, the tidal deformability \(\Lambda\) is given by
\begin{align}
  \Lambda = \frac{2k_2}{3c^5}.
\end{align}

\begin{figure}[h!]
\centering 
\includegraphics[width=.48\textwidth]{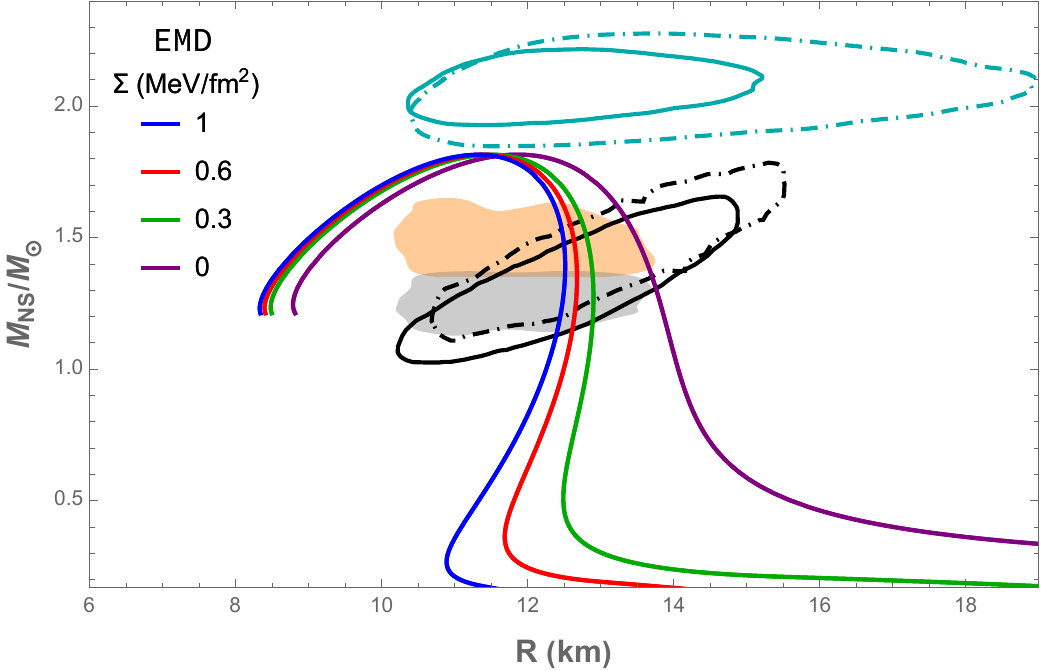}
\hfill
\includegraphics[width=.48\textwidth]{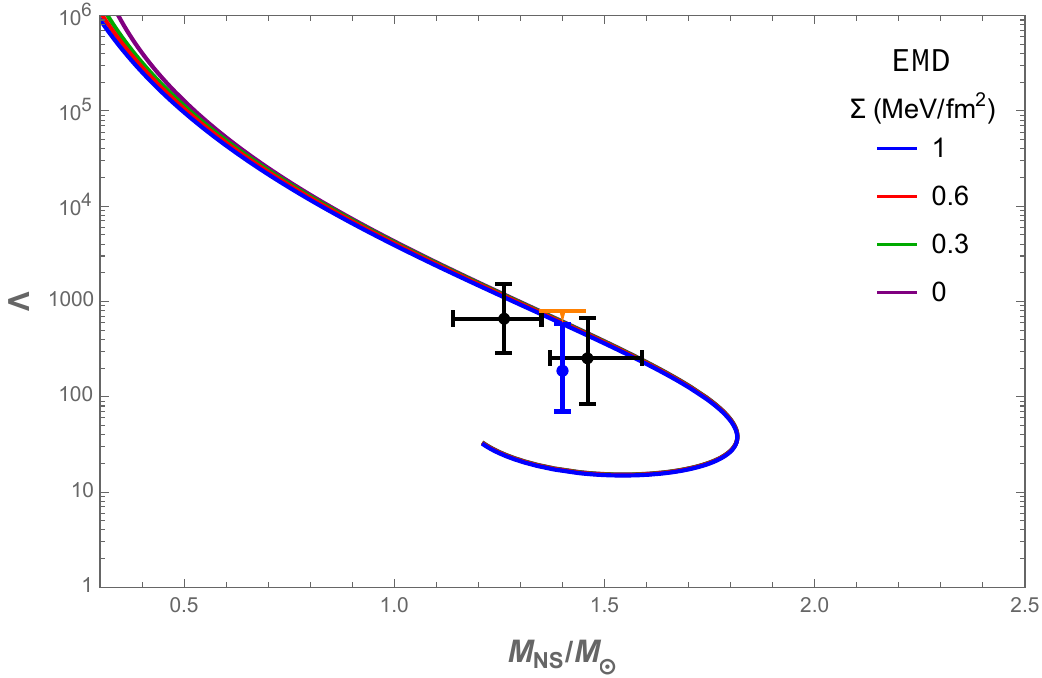}

\includegraphics[width=.48\textwidth]{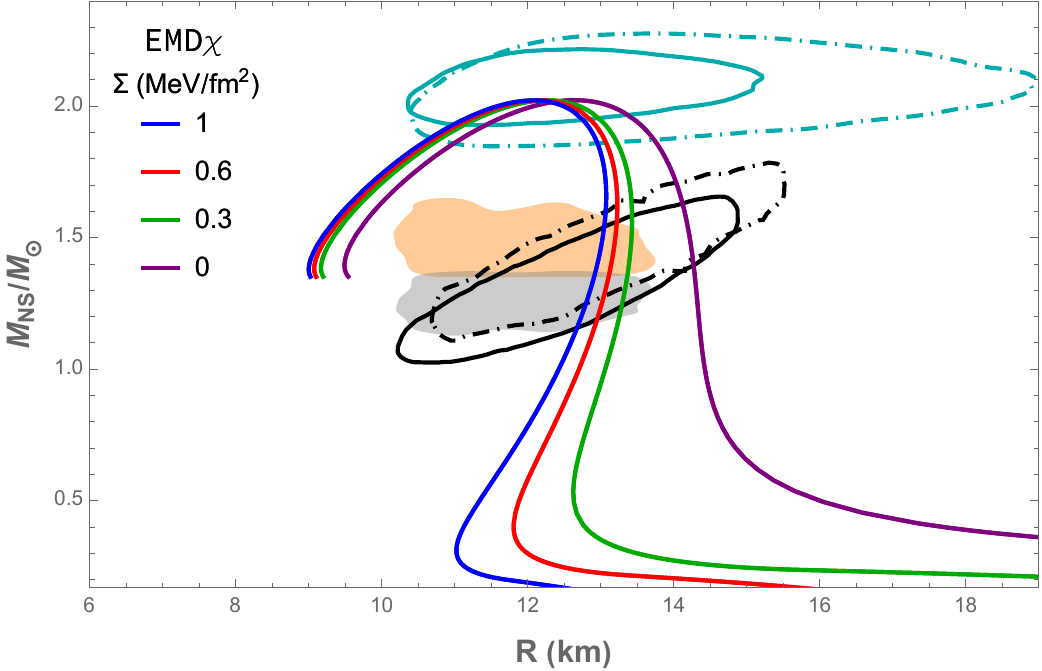}
\hfill
\includegraphics[width=.48\textwidth]{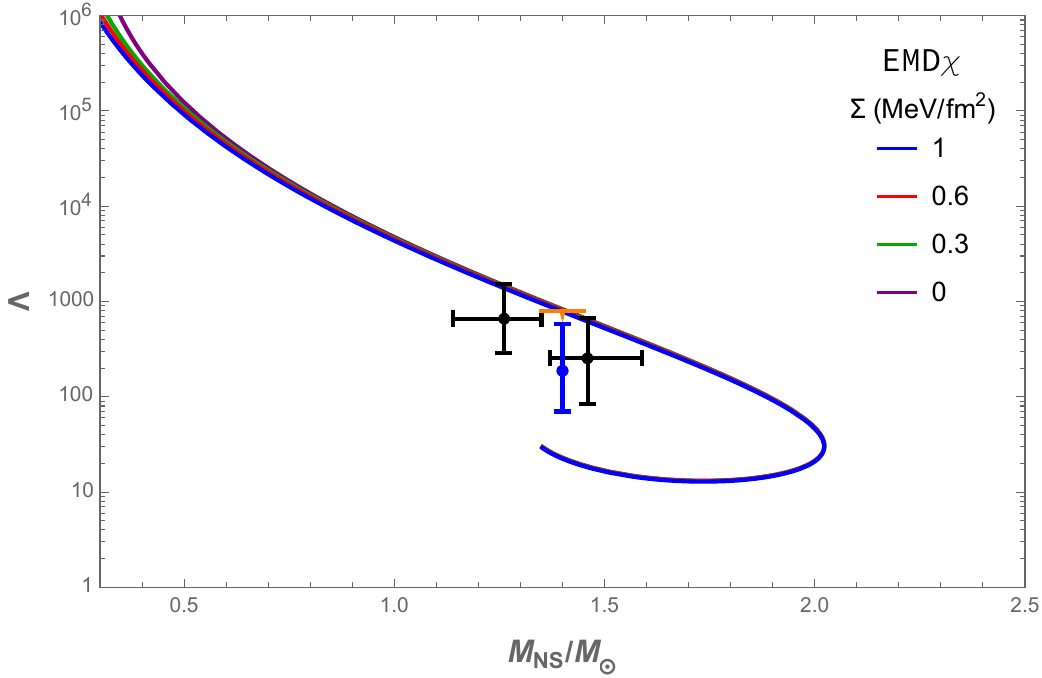}
\caption{The M-R relation and tidal deformability \(\Lambda\) of neutron stars at \(\Sigma = 0,\, 0.3,\, 0.6,\, 1 \, {\rm MeV/fm^2}\). The upper panels represent the EMD system, while the lower panels represent the EMD\(\chi\) system. In the left panels, the orange and gray shaded regions indicate constraints from the GW170817 binary neutron star merger event \cite{LIGOScientific:2017vwq, LIGOScientific:2018cki}. The black solid and dashed lines denote constraints from PSR \({\rm J}0030+0451\) \cite{Riley:2019yda, Miller:2019cac}, while the cyan solid and dashed lines indicate constraints from PSR \({\rm J}0740+6620\) \cite{Riley:2021pdl, Miller:2021qha}. Tidal deformability constraints in the right panels are derived from GW170817 \cite{LIGOScientific:2017vwq, LIGOScientific:2018cki, Fasano:2019zwm}.}
\label{NeutronStarMassRadiusAndTidalDeformability}
\end{figure}

The M-R relation and tidal deformability \(\Lambda\) have been calculated at four different surface tensions \(\Sigma\) for both the EMD and EMD\(\chi\) systems, as shown in Fig.~\ref{NeutronStarMassRadiusAndTidalDeformability}. The empirical value of \(\Sigma\) for symmetric nuclear matter at saturation is approximately \(\Sigma \simeq 1 \, {\rm MeV/fm^2}\) \cite{Hua:2000gd, Drews:2013hha}, and it is expected to decrease with increasing neutron excess \cite{Douchin:1998fe}. Given that \( n_P / n_B = 1/4 \) is adopted in this study, we examine four values of \(\Sigma\), ranging from \(0\) to \(1 \, {\rm MeV/fm^2}\). The numerical results for the maximum mass and corresponding radius of the neutron star at these values of \(\Sigma\) are presented in table~\ref{MaximumMassAndRadius}.

Constraints on the M-R relation and tidal deformability from gravitational wave and astrophysical observations are also presented in Fig.~\ref{NeutronStarMassRadiusAndTidalDeformability}. The orange and gray regions indicate constraints from the GW170817 binary neutron star merger event using the spectral EoS approach \cite{LIGOScientific:2017vwq, LIGOScientific:2018cki}. The black solid and dashed lines represent constraints from PSR \({\rm J}0030+0451\) \cite{Riley:2019yda, Miller:2019cac}, while the cyan solid and dashed lines indicate constraints from PSR \({\rm J}0740+6620\) \cite{Riley:2021pdl, Miller:2021qha}. All constraints are shown at the 90\% confidence level. The tidal deformability constraints in the figure are also based on GW170817 data. In Ref.~\cite{LIGOScientific:2017vwq}, assuming a common EoS for both neutron stars, an upper bound on the tidal deformability of 1.4 solar-mass neutron stars was determined as \(\Lambda_{1.4} \leq 800\); this upper bound is represented by the orange line in Fig.~\ref{NeutronStarMassRadiusAndTidalDeformability}. Ref.~\cite{LIGOScientific:2018cki} provides an improved estimate, \(\Lambda_{1.4} = 190^{+390}_{-120}\), shown by the blue error bar. The black error bars correspond to additional analyses of the same event \cite{Fasano:2019zwm}.

\begin{table}[H]
\fontsize{6.5pt}{10pt}\selectfont
\centering
\begin{tabular}{|c|cccc|}
\hline
\(\Sigma\,{\rm \left(MeV/fm^2\right)}\)& \(0\)& \(0.3\)&\(0.6\)&\(1\)\\
\hline 
EMD& (11.88 km,\ 1.8161 \(\rm M_\odot\))& (11.52 km,\ 1.8159 \(\rm M_\odot\))&(11.42 km,\ 1.8157 \(\rm M_\odot\))&(11.34 km,\ 1.8156 \(\rm M_\odot\))\\
EMD$\chi$& (12.64 km,\ 2.023 \(\rm M_\odot\))& (12.28 km,\ 2.0226 \(\rm M_\odot\))&(12.17 km,\ 2.0225 \(\rm M_\odot\))&(12.08 km,\ 2.0223 \(\rm M_\odot\))\\
\hline
\end{tabular}
\caption{Numerical results for the maximum mass and corresponding radius of the neutron star at surface tensions \(\Sigma = 0,\, 0.3,\, 0.6,\, 1 \, {\rm MeV/fm^2}\). The upper row represents the EMD system, while the lower row corresponds to the EMD\(\chi\) system.}
\label{MaximumMassAndRadius}
\end{table}

From Fig.~\ref{NeutronStarMassRadiusAndTidalDeformability}, we observe that the surface and Coulomb effects in the crust significantly influence the M-R relation but have minimal impact on the maximum mass of the neutron star. Moreover, a nonzero surface tension, \(\Sigma\), is necessary to achieve an M-R relation that aligns with current observational constraints. When comparing the results between the two systems, we find that the EMD\(\chi\) system supports a neutron star with a mass exceeding two solar masses, consistent with recent observational and theoretical expectations \cite{Antoniadis:2013pzd, Fonseca:2021wxt}, which require support for neutron stars of \(2.01 \pm 0.04\) and \(2.08 \pm 0.07\) solar masses. In contrast, the EMD system supports a maximum mass of approximately 1.816 solar masses, as shown in Table~\ref{MaximumMassAndRadius}. For tidal deformability \(\Lambda\), we note that the values from the EMD\(\chi\) system are slightly higher than those from the EMD system and approach the upper limits of the observational constraints in Refs.~\cite{LIGOScientific:2017vwq, Fasano:2019zwm}. The surface tension \(\Sigma\) has minimal effect on \(\Lambda\) within the region constrained by observations.

\begin{figure}[h!]
\centering 
\includegraphics[width=.48\textwidth]{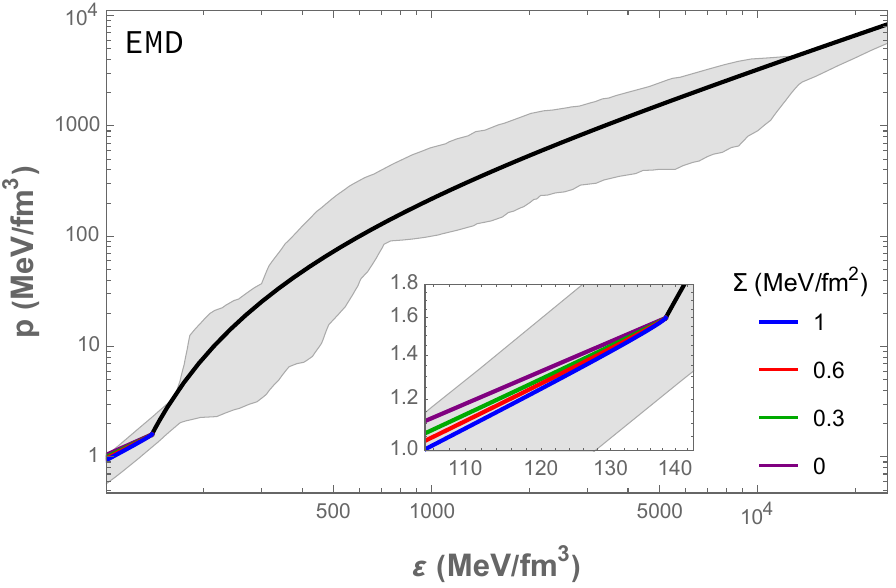}
\hfill
\includegraphics[width=.48\textwidth]{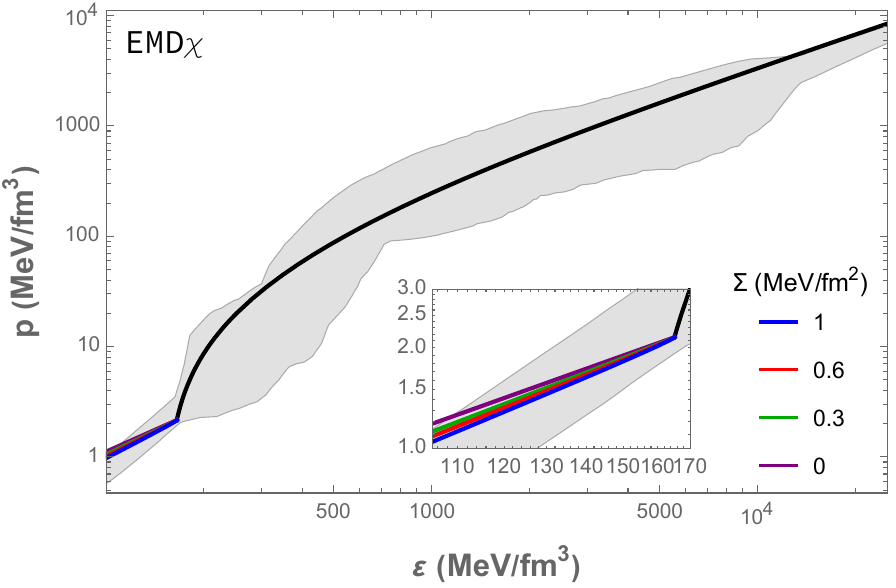}
\caption{Equations of state for the neutron star at surface tensions \(\Sigma = 0,\, 0.3,\, 0.6,\, 1 \, {\rm MeV/fm^2}\). The left panel represents the EMD system, while the right panel corresponds to the EMD\(\chi\) system. The inset images provide a magnified view of the low-energy density region.}
\label{NeutronStarPEFig-Sigma}
\end{figure}

The EoS of the neutron star at surface tensions \(\Sigma = 0,\, 0.3,\, 0.6,\, 1 \, {\rm MeV/fm^2}\) for both the EMD and EMD\(\chi\) systems are shown in Fig.~\ref{NeutronStarPEFig-Sigma}. In these $p$-$\varepsilon$ curves, the black segments represent the core of the neutron star, while the colored segments correspond to the crust at various surface tensions \(\Sigma\). We find that the holographic model developed here provides a consistent description of the neutron star EoS, aligning well with current observational and theoretical constraints. Additionally, we observe that \(\Sigma\) has minimal influence on the crust EoS within the energy density range shown in Fig.~\ref{NeutronStarPEFig-Sigma}.

\begin{figure}[h!]
\centering 
\includegraphics[width=.49\textwidth]{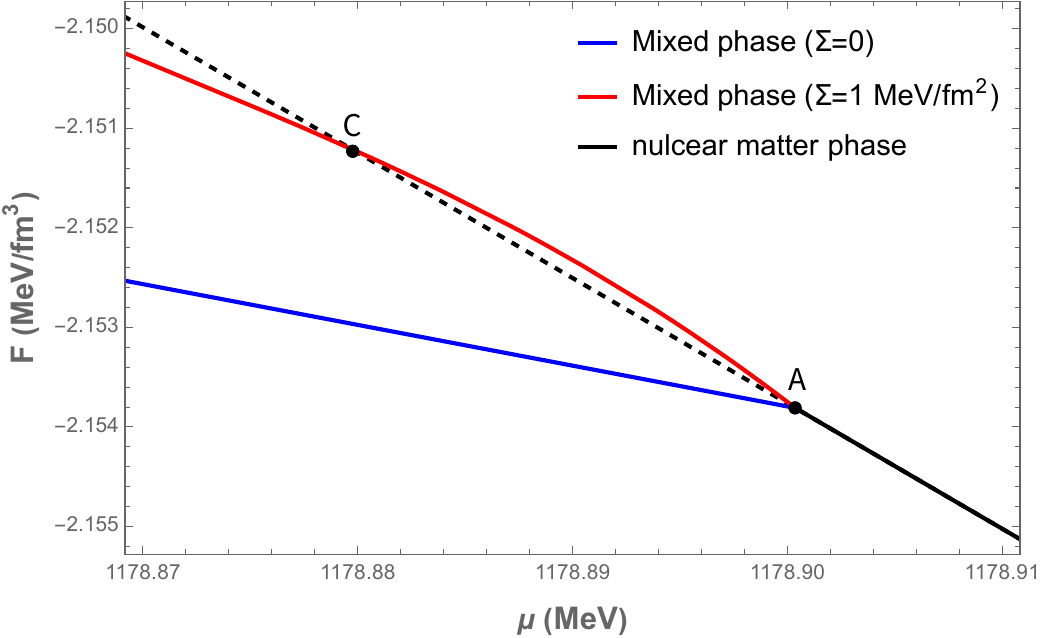}
\hfill
\includegraphics[width=.48\textwidth]{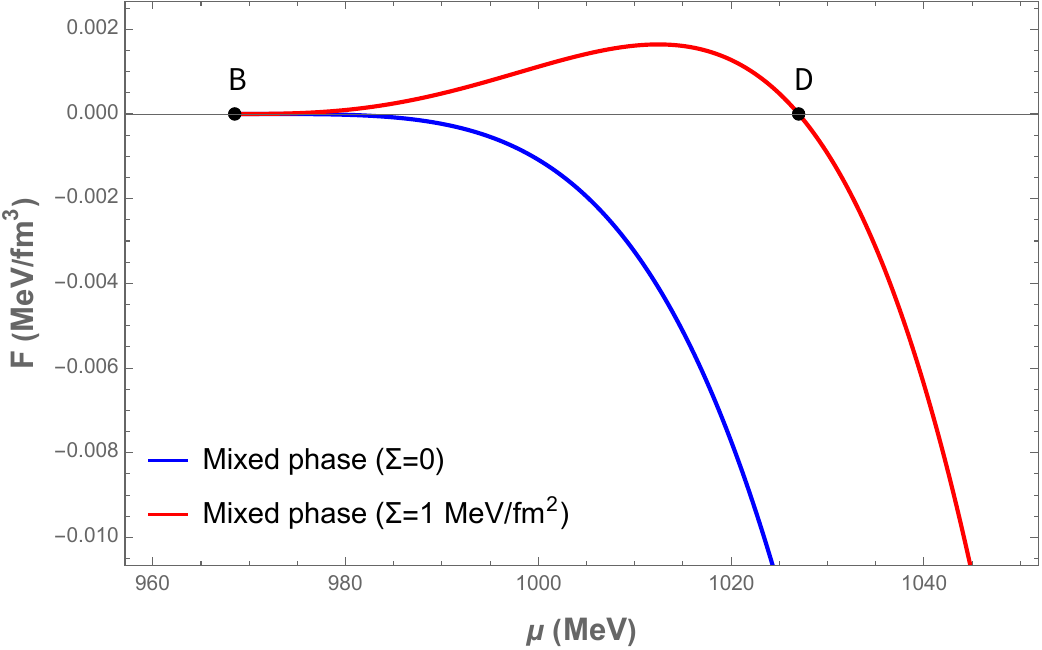}
\caption{Variations of free energy $F$ with chemical potential $\mu$ in the core-crust and crust-vacuum transition regions. The left panel illustrates the transition from the nuclear matter phase to the mixed phase, while the right panel depicts the transition from the mixed phase to the vacuum.}
\label{PhaseTransitionSchematicDiagram}
\end{figure}

It is important to note that without surface and Coulomb effects, the transitions between the core and crust, as well as between the crust and vacuum, are continuous. When these effects are included, i.e., for \(\Sigma \neq 0\), these transitions become first-order, resulting in discontinuities in energy density. This is illustrated in Fig.~\ref{PhaseTransitionSchematicDiagram}, where the blue line represents the mixed phase of the crust at \(\Sigma=0\), and the black line denotes the nuclear matter phase. 
Points A and B in Fig.~\ref{PhaseTransitionSchematicDiagram} correspond to the original transition points between the core and crust, and between the crust and vacuum, respectively. When surface and Coulomb effects are considered, the blue line shifts to the red line, as described by Eq.~(\ref{CrustTotalPressureWithSurface}). This shift introduces two new transition points, labeled C and D, where the free energy in segments AC and BD is lower than in the mixed phase. These new transition points introduce discontinuities in both energy density and particle number density. In our work, however, the first-order transition between the core and crust remains very weak, closely resembling a continuous one. This contrasts with the findings in Ref.~\cite{Kovensky:2021kzl}, where a distinct discontinuity is present at the transition.

\section{Conclusion and discussion}\label{ConclusionSection}

In this work, we analyze a specific EMD$\chi$ system and an EMD system with the coupling set to $\beta = 0$, both of which can model QCD phase transitions. By fixing the model parameters, we are able to reproduce the behavior of thermodynamic quantities, including pressure, energy density, and baryon number susceptibility as functions of temperature, in agreement with lattice QCD results \cite{Karsch:2001vs,Datta:2016ukp}. We then extend our investigation to finite chemical potentials, examining phase transition properties and presenting the QCD phase diagram. Both models successfully capture transition behaviors that closely align with lattice results \cite{Datta:2016ukp}. The CEP in our models is located at $\left(\mu_{\rm CEP},T_{\rm CEP}\right) =\left(233 \, \text{MeV}, 149.8 \, \text{MeV}\right)$ for the EMD system and $\left(\mu_{\rm CEP},T_{\rm CEP}\right) = \left(284 \, \text{MeV}, 138.6 \, \text{MeV}\right)$ for the EMD$\chi$ system. Additionally, we observe a first-order phase transition at $\mu_B = 0$ in the gauge sector of the EMD$\chi$ system, consistent with results from pure gauge theory \cite{Boyd:1996bx,Fukushima:2010bq}.

We also investigate the low-temperature and high-density conditions relevant to neutron stars within our models. In constructing the neutron star, we apply these bottom-up holographic models to describe baryonic matter, with the lepton component introduced to maintain charge neutrality. Our holographic model for neutron stars includes both a core and a crust structure, with surface and Coulomb effects incorporated. The neutron star EoS, as presented in Fig.~\ref{NeutronStarPEFig-Sigma}, demonstrates improved low-energy behavior within the chiral effective theory region and enhanced high-energy behavior in the perturbative QCD region compared to previous studies \cite{Kovensky:2021kzl}. Our EoS remains within the allowed range specified in Ref.~\cite{Annala:2019puf} and aligns with astrophysical constraints \cite{LIGOScientific:2017vwq, Riley:2019yda, Miller:2019cac, Riley:2021pdl, Miller:2021qha, Fasano:2019zwm, Antoniadis:2013pzd, NANOGrav:2019jur, Fonseca:2021wxt}. Furthermore, the maximum neutron star mass predicted by the EMD$\chi$ system exceeds 2 solar masses, consistent with recent observational findings \cite{Antoniadis:2013pzd,Fonseca:2021wxt}.

In future work, several interesting extensions to the models considered here could be explored. Firstly, while the present model is based on a two-flavor system, an extension to a 2+1 flavor system could be achieved by introducing an additional vacuum scalar field associated with the strange quark flavor. This extension would allow for comparisons between $ud$ neutron stars and $uds$ neutron stars, as examined in previous studies \cite{Holdom:2017gdc, Yuan:2022dxb}. Furthermore, we could investigate the effects of temperature-dependent neutron star EOS and magnetic fields \cite{Gursoy:2017wzz, Cai:2024eqa, Arefeva:2022avn}. Including these factors in holographic models and studying their impacts on neutron stars is relatively straightforward yet highly valuable, particularly in the context of supernova explosions neutron star mergers \cite{Baiotti:2016qnr}. This would also enable a more complete treatment of phase transitions within neutron stars, potentially facilitating models of heavier neutron stars \cite{Tan:2021ahl} and yielding improvements in predictions for tidal deformability \(\Lambda\) \cite{Han:2018mtj}. With the rise of multi-messenger astronomy, research on neutron stars and other compact objects has gained significant importance. These compact stars serve as natural laboratories—unachievable on Earth—for studying the properties of extremely dense matter, and offer unique insights that can greatly advance our understanding of fundamental physics.

\section*{Acknowledgments}
We would like to thank Xiao-Chang Peng and Prof. Jinniu Hu for their helpful comments and discussions. This work is supported by Hunan Provincial Natural Science Foundation of China (Grant Nos. 2023JJ30115 and 2024JJ3004), by the YueLuShan Center for Industrial Innovation (2024YCII0117), and by the National Key Research and Development Program of China under Grant No.2020YFC2201501, and by the National Science Foundation of China (NSFC) under Grants No.~12347103 and No.~11821505. 

\bibliographystyle{unsrt}
\bibliography{refs-NeutronStar.bib}

\end{document}